\documentclass[twocolumn,preprintnumbers,prb,reprint,aps,twocolumn,pre,nofootinbib,longbibliography]{revtex4-1}

\usepackage{amsmath,amssymb,amsfonts,amsbsy,amsthm}
\usepackage{xcolor}
\usepackage{epsfig}
\usepackage{graphicx,placeins}
\usepackage{bm}

\newcommand{\e}{{\bm e}}
\newcommand{\ii}{\rm i}
\newcommand{\tn}{{\tilde n}}
\newcommand{\tm}{{\tilde m}}

\newcommand{\E}{{\bm E}}

\newcommand{\m}{{\bm m}}
\newcommand{\p}{{\bm p}}

\renewcommand{\j}{{\bm j}}
\renewcommand{\k}{{\bm{k}}}
\newcommand{\q}{{\bm{q}}}
\renewcommand{\r}{{\bm{r}}}

\renewcommand{\Re}{{\rm{Re}}}
\renewcommand{\Im}{{\rm{Im}}}

\usepackage[normalem]{ulem}

\def\XXint#1#2#3{{\setbox0=\hbox{$#1{#2#3}{\int}$ }
\vcenter{\hbox{$#2#3$ }}\kern-.6\wd0}}

\def\gsim{\lower.35em\hbox{$\stackrel{\textstyle>}{\textstyle\sim}$}}

\begin{document}
\title{Neutral magic-angle bilayer graphene: Condon instability and chiral resonances}

\author{T. Stauber$^{1,2}$, M. Wackerl$^{2}$, P. Wenk$^{2}$, D. Margetis$^{3}$, J. Gonz\'alez$^{4}$, G. G\'omez-Santos$^{5}$, and J. Schliemann$^{2}$}

\affiliation{$^{1}$Instituto de Ciencia de Materiales de Madrid, CSIC, E-28049 Madrid, Spain\\
$^{2}$Institut f\"ur Theoretische Physik, Universit\"at Regensburg, Germany\\
$^{3}$Institute for Physical Science and Technology, and Department of Mathematics, and Center for Scientific Computation and Mathematical Modeling, University of Maryland, College Park, Maryland 20742, USA\\
$^{4}$Instituto de Estructura de la Materia, CSIC, E-28006 Madrid, Spain\\
$^5$Departamento de F\'{\i}sica de la Materia Condensada, Instituto Nicol\'as Cabrera and Condensed Matter Physics Center (IFIMAC), Universidad Aut\'onoma de Madrid, E-28049 Madrid, Spain}
\date{\today}

\begin{abstract}
We discuss the full optical response of twisted bilayer graphene at the neutrality point close to the magic angle within the continuum model (CM). Firstly, we identify three different channels consistent with the underlying $D_3$ symmetry, yielding the total, magnetic, and chiral response. Secondly, we numerically calculate the full optical response in the immediate vicinity of the magic angle $\theta_m$ which provides a direct mapping of the CM onto an effective two-band model. We, further, show that the ground-state of the CM in the immediate vicinity of $\theta_m$ is unstable towards transverse current fluctuations, a so-called Condon instability. Thirdly, due to the large counterflow, the acoustic plasmonic excitations with typical wave numbers have larger energies than the optical ones and their energy density may be largely enhanced at certain frequencies which we denominate as {\it chiral resonances}. Finally, we discuss symmetry relations for the optical response and their consequences for the chiral response.
\end{abstract}

\maketitle
\section{Introduction}
Twisted bilayer graphene\cite{Lopes07,Shallcross08,Suarez10,Schmidt10,Li10,Trambly10,Bistritzer11,Lopes12,Dean13} has attracted much attention due to a plethora of new correlated phases such as correlated insulators,\cite{Cao18a} unconventional superconductivity\cite{Cao18b,Yankowitz18} or anomalous quantum Hall ferromagnetism.\cite{Sharpe19,Polshyn20} Most features are related to the emergence of a flat band which is related to the vanishing of the Fermi velocity at the so-called magic angle $\theta_m$. In addition, a prominent counterflow can be found where the current has opposite direction with respect to the two layers and which becomes balanced at $\theta_m$.\cite{Bistritzer11} A third feature is the pronounced circular dichroism\cite{Kim16,Suarez17} at frequencies close to the van Hove singularities.  

Fermi velocity, counterflow and circular dichroism are related to the total (electric), magnetic and chiral response. These were first introduced in Refs. \onlinecite{Stauber18,Stauber18b} and discussed in detail for large twist angles. Here, we shall calculate these quantities for twist angles around the magic angle. 

We will also address flat band plasmonics in twisted bilayer graphene\cite{Hu17} that is related to localized collective modes.\cite{Stauber16} This topic is an area of active interest.\cite{Levitov19,Novelli20,Hesp21,Huang22,Kuang21} Finally, we give a general discussion and new insights on the chiral optical response and how it is related to symmetries. 

The flat bands\cite{Bistritzer11,Tarnopolsky19,SongZhida19,Hejazi19,Koshino19,Park20} and the optical response\cite{Stauber13,Vela18,Novelli20,Dai21,Han22} of twisted bilayer graphene have been investigated in numerous articles so far. However, a detailed discussion on the scaling behavior of the response function for small frequencies, as $\omega\to0$, around the immediate vicinity of the magic angle, is missing until now. Another topic is related to the Condon instability\cite{Condon68} that has recently been discussed in related systems. \cite{Andolina20,Nataf19,Guerci20,Sanchez21} We will argue that the Condon instability arises in the continuum model\cite{Lopes07,Bistritzer11} of twisted bilayer graphene at the neutrality point in the immediate vicinity of the magic angle. 
 
Apart from the above, the large counterflow or magnetic response has also been discussed in several papers.\cite{Bistritzer11,Stauber18,Guerci21} Nevertheless, the importance of {\it acoustic} plasmonic modes has not received sufficient attention, so far. We believe that our results will be relevant to flat-band plasmonics in twisted bilayer graphene, especially at the {\it chiral resonance} where the energy density can be largely enhanced.

Finally, the chirality in graphene might be used to design novel cavities that lead to strong chiral light-matter interaction.\cite{Stauber20} In order to understand the underlying physics we point out some new aspects related to particle-hole symmetry. This leads us to distinguish between electron and hole transitions where the initial states are energetically closer and further away from the neutrality point than the final states, respectively. 

The paper is organized as follows. In Sec. II, we define the continuum model for twisted bilayer graphene. In Sec. III, we introduce the minimal model for the full linear response that defines the total, magnetic, and chiral responses. In Sec. IV, we present our numerical results for the dissipative and reactive response. In Sec. V, we carry out this task, albeit in the immediate vicinity of the magic angle. By this procedure, we obtain a scaling relation that eventually leads to the prediction of a Condon instability. In Sec. VI, we discuss flat-band plasmonics without excess charges for twist angles near the magic angle, and highlight the fact that genuinely acoustic plasmons might be dominating the plasmonic properties around the magic angle. In Sec. VII, we outline the symmetry conditions for chiral response. Finally, we conclude our paper with Sec. VIII, a summary and outlook. Supplementary Information on the numerical recipe of how to calculate the dissipative response in the clean limit as well as on an analytical calculation of the optical conductivity in the immediate vicinity of the magic angle is also provided.
\section{Hamiltonian}
The local Hamiltonian of a twisted bilayer graphene can be approximated by\cite{Moon13}
\begin{align}
\label{HamiltonianSpace}
H=\left(\begin{matrix}H_0^{-\theta/2}&V^\dagger(\r)\\V(\r)&H_0^{\theta/2}\end{matrix}\right)\;,
\end{align}
where $H_0^\gamma=-i\hbar v_F{\bm \tau}^\gamma\cdot\partial_\r$ denotes the Hamiltonian of the separate layers with  $( \bm \tau^{\gamma}_x,\bm \tau^{\gamma}_y ) =  e^{\ii\gamma\bm\tau_z/2} ( \bm \tau_x, \bm \tau_y ) e^{-\ii\gamma\bm\tau_z/2}  $, $\bm \tau_{x,y,z}$ being the Pauli matrices. The interlayer coupling $V(\r)$ also denotes a $2\times2$-matrix and defines the coupling between the layers. It explicitly depends on the stacking order, but a common approximation is that all components are defined by only one common function $u(\r)$.\cite{Lopes07,Moon13} Expanding $u(\r)$ into the first three Fourier-components of the moir\'e lattice and representing the Hamilton operator by plane waves, one arrives at the non-interacting Hamiltonian used for calculating the response to total fields\cite{Lopes07,Bistritzer11}
\begin{align}\label{Hamiltonian}
\mathcal{H } =&\hbar v_F\sum_{\bm k;\alpha,\beta} c_{\bm k,\alpha,1}^{\dagger} \;\bm \tau_{\alpha\beta}^{-\theta/2} \cdot \bm k \; c_{\bm k,\beta,1} \notag\\ 
+&\hbar v_F\sum_{\bm k;\alpha,\beta} c_{\bm k,\alpha,2}^{\dagger} \;\bm \tau_{\alpha\beta}^{+\theta/2} \cdot \bm k \; c_{\bm k,\beta,2}\notag\\
+&\frac{t_{\perp}}{3}\sum_{\bm k;\alpha,\beta;\bm G} (c_{\bm k + \bm G,\alpha,1}^{\dagger} \; T_{\alpha\beta}(\bm G) \; c_{\bm k,\beta,2} + H. c.)\;,
\end{align}
where the separation between twisted cones is $\Delta \bm K = 2 |\bm K| \sin(\theta/2)  \left(0,1\right)$  with $\bm K = \tfrac{4 \pi}{3 a_g} \left(1,0\right)$. Interlayer hopping is restricted to wavevectors $\bm G=\bm 0,\bm G_1,\bm G_2$ with $\bm G_1 =  |\Delta \bm K| \left( \tfrac{-\sqrt{3}}{2},-\tfrac{3}{2}\right)$, $\bm G_2 =  |\Delta \bm K| \left(\tfrac{\sqrt{3}}{2},-\tfrac{3}{2}\right)$, and
\begin{equation}\label{Hopping}
T(\bm 0) = \left(\begin{matrix} \kappa & 1\\1 & \kappa\end{matrix}\right); \; \; T(\bm G_1) = T^{*}(\bm G_2)=\left(\begin{matrix}
\kappa e^{\ii 2 \pi / 3} & 1\\ e^{-\ii 2 \pi / 3}& \kappa e^{\ii 2 \pi / 3}\end{matrix}\right)
.\end{equation}
Calculations are performed with $t=2.78\,\text{eV}$ and $t_{\perp}=0.33\,\text{eV}$, being $\hbar v_F= \tfrac{\sqrt{3}}{2} t a_g $ the Fermi velocity with
 graphene lattice constant
$a_g=2.46 \, \mathring{\text{A}}$; the interlayer distance has been taken as $a=3.5 \, \mathring{\text{A}}$. In the first part of the work, we discuss the symmetric model with $\kappa=1$ and in the second part of the work, the asymmetric model introduced in Ref. \onlinecite{Koshino19} with $\kappa=0.8$ is used that accounts for out-of-plane relaxation, see also Ref. \onlinecite{Guinea19}. 

Let us finally note that besides the parameter $\kappa$, the above model is only characterized by one dimensionless parameter $\alpha_{\theta_i}=\frac{\sqrt{A_i/3}}{2\pi}\frac{t_\perp}{t}$ which combines $t_\perp$ and the twist angle $\theta_i$ parametrized by $i$ via $A_i=3i^2 + 3i + 1$ with $\cos(\theta_i)=1-\tfrac{1}{2A_i}$.\cite{Bistritzer11} This can readily be seen from the Hamiltonian of Eq. (\ref{HamiltonianSpace}) by introducing the dimensionless coordinates $\tilde\r=|\Delta \bm K|\r$ such that the new interlayer coupling between the layers is independent of the twist angle.\cite{Lopes07} In principle, $i\in\mathbb{N}$ denotes a commensurate twist angle, but the expressions can be generalised to arbitrary real numbers $i\in\mathbb{R}$.\cite{Bistritzer11} 
\section{Linear response}
To describe chiral effects without breaking time-reversal or rotational ($C_3$) symmetry, we have to treat an effectively three-dimensional system. The minimal model thus consists of treating each layer of the twisted bilayer separately. With the Kubo formula $j_\alpha^\ell=-\chi_{j^{\ell}_\alpha\, j^{\ell'}_\beta}A_\beta^{\ell'}$, where $A_\alpha^\ell$ denotes the gauge field and summation over repeated indices is implied, the $4\times4$ local ($\q=0$) conductivity tensor then is 
\begin{equation}
\sigma^{\ell,\ell'}_{\alpha\beta}(\omega)= i \frac{\chi_{j^{\ell}_\alpha\, j^{\ell'}_\beta}(\omega)}{\omega+i0^+} 
\label{KuboConductivity}
,\end{equation}
with axis indices $\alpha,\beta=x,y $ and plane indices $\ell,\ell'=1,2$. 

The retarded current-current response is given by 
\begin{widetext}
\begin{equation}\label{KuboCurrent}
\chi_{j^{\ell}_\alpha\, j^{\ell'}_\beta}(\omega) = g_s g_v\int_{1.BZ} \frac{d^2 \bm k}{(2 \pi)^2} \sum_{n,m}
\frac{n_F(\epsilon_{m,\bm k}) - n_F(\epsilon_{n,\bm k})}{\hbar\omega+i0^+ - \epsilon_{n,\bm k} + \epsilon_{m,\bm k}} 
\langle m,\bm k|j^{\ell}_\alpha |n,\bm k\rangle \langle n,\bm k|j^{\ell'}_\beta |m,\bm k\rangle\;.\end{equation}
\end{widetext}
Here, $g_s=g_v=2$ are the spin and valley degeneracies. The states $|m, \bm k\rangle$ are eigenstates of $\mathcal{H}$ in subband $m$ and of momentum $\bm k$ in the first Brillouin zone of the superstructure. Their eigenenergies are $\epsilon_{n,\bm k} $ and $n_F$ is the Fermi function. For single layer graphene the current operator is $\bm j =-e  v_F\bm \tau$ and also for twisted bilayer graphene with the Hamiltonian of Eq. (\ref{Hamiltonian}), the general current operator is independent of $\k$.

The full current response due to an applied in-plane electric field that satisfies rotational (or $C_3$) and time-reversal symmetry reads 
\begin{equation}\label{CurrentSymmetries}
\left(\begin{matrix} j_x^1\\j_y^1\\j_x^2\\j_y^2\end{matrix}\right)= \left(\begin{matrix} \sigma_0 & 0&\sigma_1&\sigma_{xy}\\0&\sigma_0&-\sigma_{xy}&\sigma_1\\\sigma_1&-\sigma_{xy}&\sigma_0&0\\\sigma_{xy}&\sigma_1&0&\sigma_0\end{matrix}\right)\left(\begin{matrix} E_x^1\\E_y^1\\E_x^2\\E_y^2\end{matrix}\right)\;.
\end{equation}
The conductivities $\sigma_\mu=i \frac{\chi_\mu(\omega)}{\omega+i0^+}$ with $\mu=0,1,xy$ are defined via the following current-current response functions:
\begin{align}
\chi_0&=\chi_{j^{1}_x, j^{1}_x}=\chi_{j^{2}_y, j^{2}_y}\\
\chi_1&=\chi_{j^{1}_x, j^{2}_x}=\chi_{j^{1}_y, j^{2}_y}=\chi_{j^{2}_x, j^{1}_x}=\chi_{j^{2}_y, j^{1}_y}\\
\chi_{xy}&=\chi_{j^{1}_x, j^{2}_y}=-\chi_{j^{1}_y, j^{2}_x}=-\chi_{j^{2}_x, j^{1}_y}=\chi_{j^{2}_y, j^{1}_x}
\end{align}
Note that the electric field may be different at the two layers and that the above symmetries allow for different in-plane conductivities in layer 1 and 2 which may arise due to a perpendicular gate voltage.\cite{Stauber18} Also the influence of a perpendicular magnetic field can be included.\cite{Margetis21} 

In the following we will discuss the response functions that transform with respect to the irreducible representations of the underlying lattice symmetry group $D_3$ which consists of two one-dimensional and one two-dimensional representation. The response functions of the total (electronic) current density $\j_{\rm{tot}}=\j^1+\j^2$ and of the magnetic current density $\j_{\rm{mag}}=\j^1-\j^2$ transform as the one-dimensional representations $A_1$ and $A_2$, respectively. These current densities are induced by an in-plane electric and magnetic field, respectively.\cite{Stauber18,Sanchez21} The chiral response involves the two current densities $j_x^1$ and $j_y^2$ which transform as the two-dimensional representation $E$. This defines the total, magnetic and chiral response, respectively, as
\begin{align}
\sigma_{\rm{tot}}&=2(\sigma_0+\sigma_1)\;,\label{SigmaTot}\\
\sigma_{\rm{mag}}&=2(\sigma_0-\sigma_1)\;,\label{SigmaMag}\\
\sigma_{\rm{\rm{chi}}}&=\sigma_{xy}\;.\label{SigmaChiral}
\end{align}
This also defines the current-current response $\chi_\nu=-i(\omega+i0^+)\sigma_\nu$ and the Drude weight $D_\nu=\lim_{\omega\to0}\omega\Im\sigma_\nu(\omega)$ with $\nu=\rm{tot},\rm{mag},\rm{\rm{chi}}$. The subindices $xy$ and $\rm{\rm{chi}}$ can be used interchangeably. 

The above definitions also allow to deduce the exact symmetry relations when the twist angle is reversed, i.e., when the opposite enantiomer is considered:\cite{Stauber18}
\begin{align}
\sigma_0(\theta)&=\sigma_0(-\theta)\\
\sigma_1(\theta)&=\sigma_1(-\theta)\\
\sigma_{xy}(\theta)&=-\sigma_{xy}(-\theta)
\end{align}
It thus suffices to only consider the response for one twist-direction.

\section{Optical response around the magic angle}
The current response consists of a dissipative (imaginary) and reactive (real) response. Numerically, the dissipative part is in principle equivalent to the evaluation of a generalized density of states. Thus, we have
\begin{widetext}
\begin{align}
\label{DissipativeResponse}
\Im\chi^{\ell,\ell'}_{\alpha\beta}(\omega) =\pi\frac{g_s g_v}{A}\sum_{\bm k}\sum_{n,m}\mathcal{O}_{m,n,{\bm k};\alpha\beta}^{\ell,\ell'}
\left[n_F(\epsilon_{m,\bm k}) - n_F(\epsilon_{n,\bm k})\right]\delta(\hbar\omega - \epsilon_{n,\bm k} + \epsilon_{m,\bm k})\;,
\end{align}
\end{widetext}
where we introduced the transition matrix element 
\begin{equation}
\mathcal{O}_{m,n,{\bm k};\alpha\beta}^{\ell,\ell'}=\langle m,\bm k|j^{\ell}_\alpha |n,\bm k\rangle\langle n,\bm k|j^{\ell'}_\beta |m,\bm k\rangle\;.
\end{equation}
Note that the matrix elements can always be considered as real since the final spectral density must be real when time-reversal symmetry is not broken (which is the case here).

Since the current response function is an analytic function in the upper $\omega$-complex plane, see Eq. (\ref{KuboCurrent}),  
the real part is obtained from the Cauchy or Kramers-Kronig relation. By calculation of the principal value integral, this procedure gives
\begin{align}
\label{KramersKronig}
\Re\chi^{\ell,\ell'}_{\alpha\beta}(\omega) =\frac{1}{\pi}\mathcal{P}\int_{-\infty}^\infty d\omega'\frac{\Im\chi^{\ell,\ell'}_{\alpha\beta}(\omega')}{\omega'-\omega}\;,
\end{align}
which can be written as an integral over only positive frequencies using $\Im\chi^{\ell,\ell'}_{\alpha\beta}(\omega)=-\Im\chi^{\ell,\ell'}_{\alpha\beta}(-\omega)$.

Due to its linear dispersion, the continuum model does not possess a nominal diamagnetic term. However, since the integral of Eq. (\ref{KramersKronig}) extends over all frequencies, one needs to invoke a high-frequency cut-off and the resulting term can be viewed as an effective diamagnetic term. For frequencies beyond the cut-off, we assume that we have the current response of decoupled graphene layers. Details of the regularization procedure can be found in Ref. \onlinecite{Stauber13}.

\begin{figure*}[t]
\includegraphics[width=0.68\columnwidth]{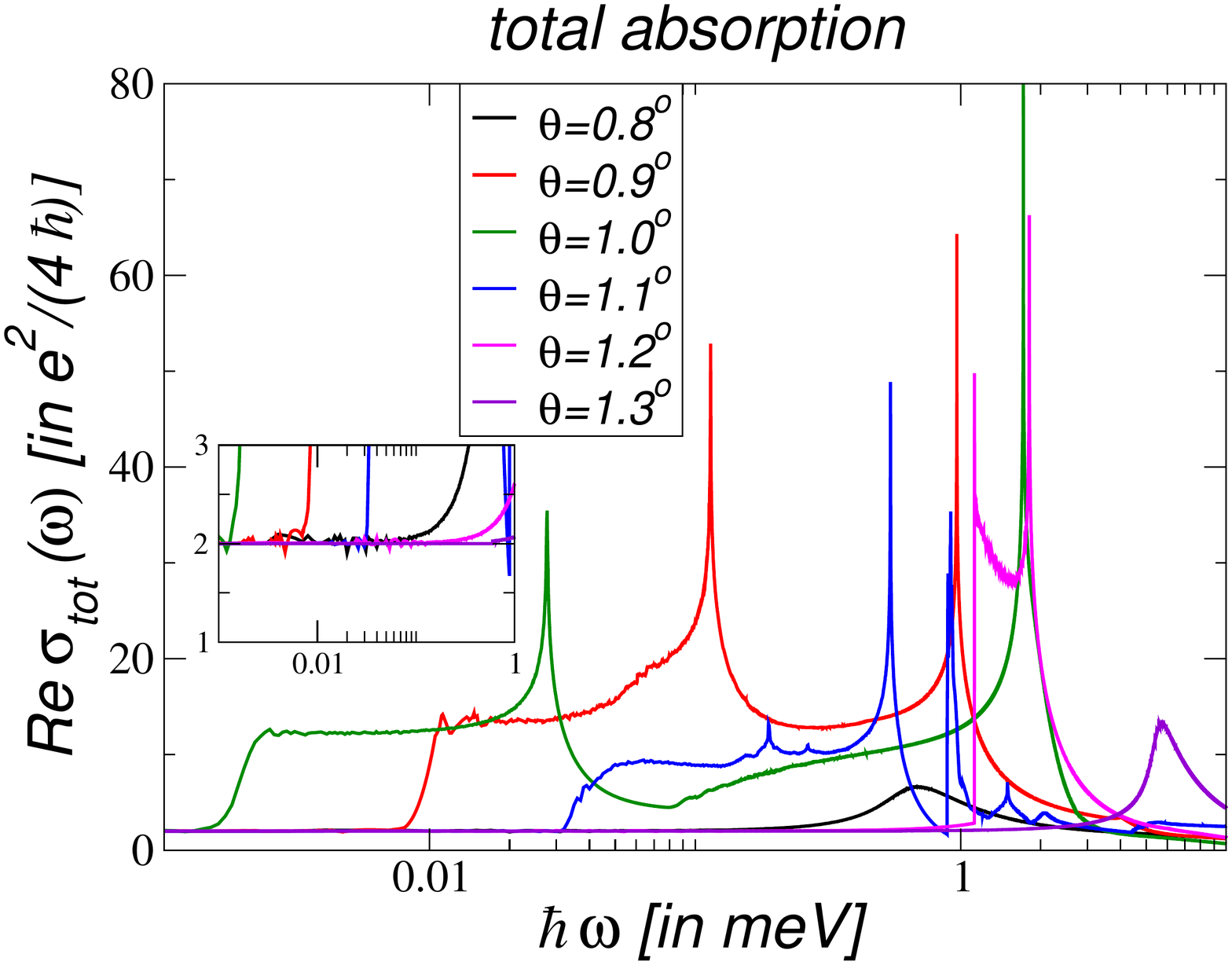}
\includegraphics[width=0.68\columnwidth]{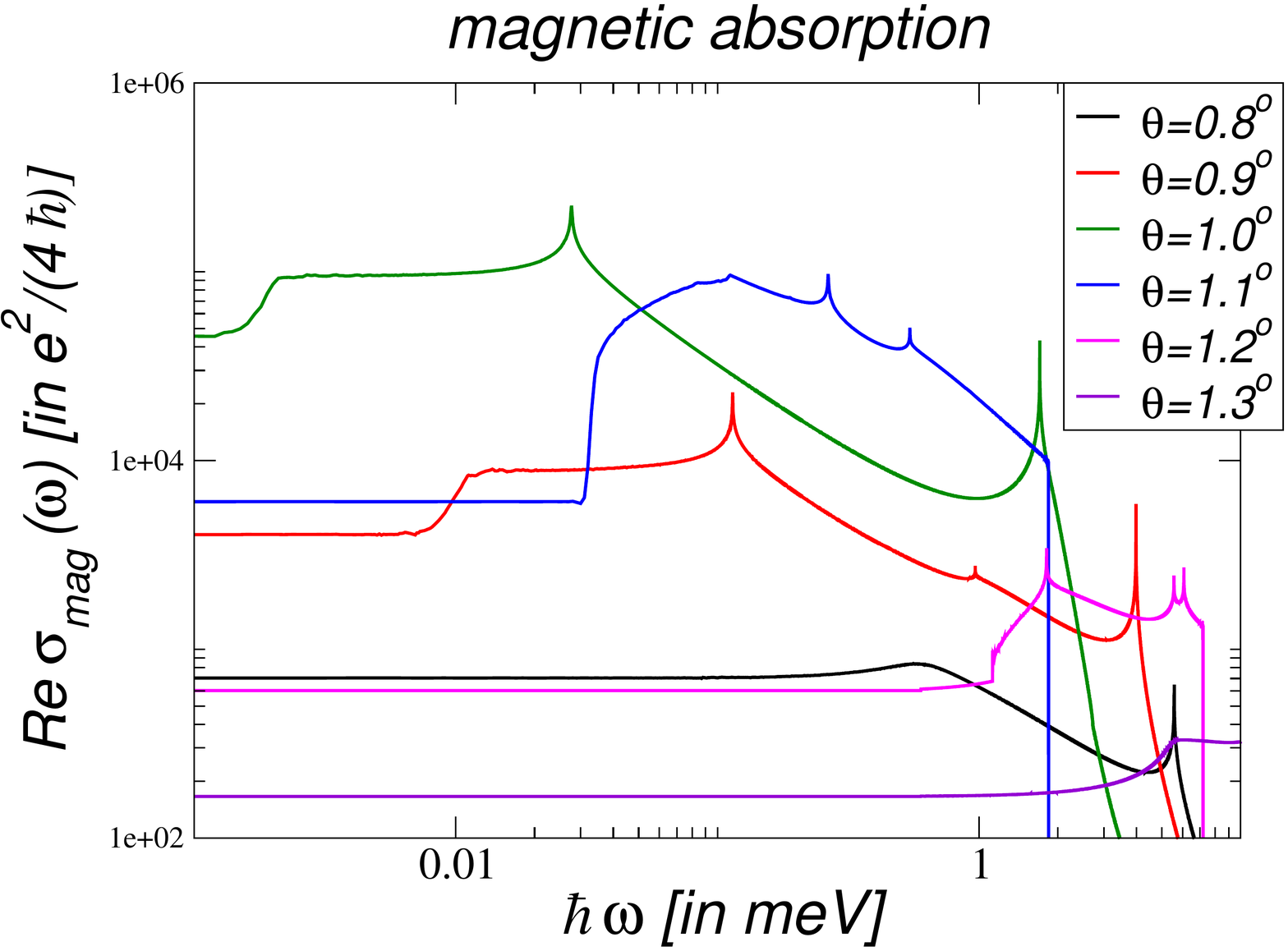}
\includegraphics[width=0.68\columnwidth]{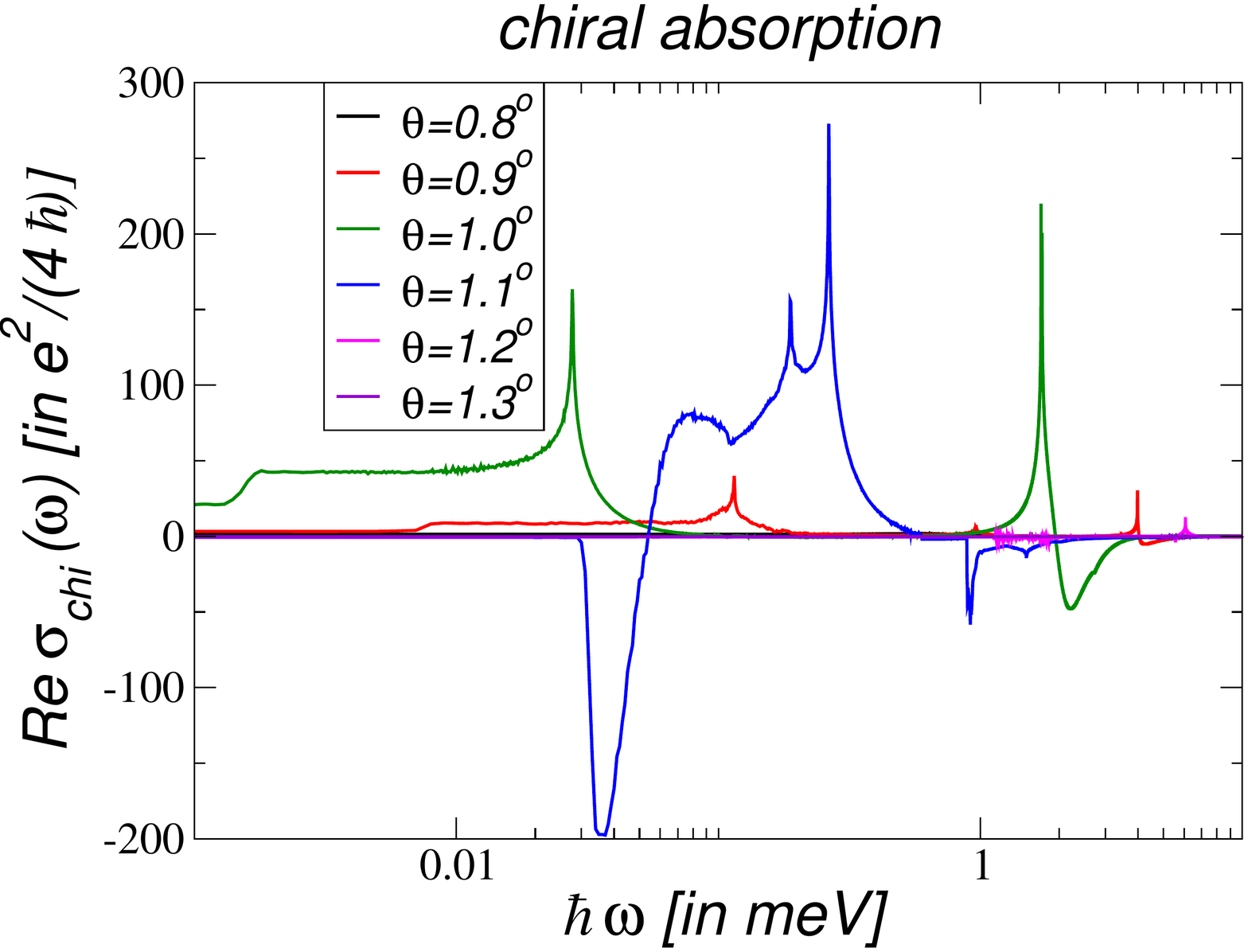}
\caption{\label{FullResponseIm} Dissipative response of the total (left), magnetic (center), and chiral (right) current for the symmetric continuum model $\kappa=1$ at various twist angles around the magic angle $\theta_m\simeq1.03^\circ$. The real  conductivities Re$\sigma_{\rm{tot}}$, Re$\sigma_{\rm{mag}}$, and Re$\sigma_{\rm{chi}}$ are given in units of the universal absorption of single-layer graphene, $\sigma_G=\frac{e^2}{4\hbar}$. The inset of the left panel shows the universal absorption of two graphene layers, $2\sigma_G$, independent of the twist angle.}
\end{figure*}

Furthermore, the complex conductivity can be obtained by first considering the dissipative (real) part and then the reactive (imaginary) part. The only difference is the Drude term that needs to be added to the dissipative part as follows:
\begin{widetext}
\begin{align}
\label{RealConductivity}
\Re\sigma_{\alpha\beta}^{\ell,\ell'}(\omega) =\pi D_{\alpha\beta}^{\ell,\ell'}\delta(\omega)+\frac{\pi}{\omega}\frac{g_s g_v}{A}\sum_{\bm k}\sum_{n,m}\mathcal{O}_{m,n,{\bm k};\alpha,\beta}^{\ell,\ell'}
\left[n_F(\epsilon_{m,\bm k}) - n_F(\epsilon_{n,\bm k})\right]\delta(\omega - \epsilon_{n,\bm k} + \epsilon_{m,\bm k})\;,
\end{align}
\end{widetext}
where
\begin{align}
D_{\alpha\beta}^{\ell,\ell'}=\lim_{\omega\to0}\chi_{j^{\ell}_\alpha\, j^{\ell'}_\beta}(\omega)
\end{align}
denotes the Drude weight matrix. According to the above equation, also the total, magnetic, and chiral Drude weight can be defined according to Eqs. (\ref{SigmaTot}-\ref{SigmaChiral}).

For the dissipative response, we will only discuss the regular term of the real conductivities which is characterized by plateaus in the low-frequency limit $\omega\to0^+$, i.e., we neglect the Drude weight which only contributes for $\omega=0$ in the absence of intrinsic damping. These plateaus are denoted as 
\begin{align}
\sigma_{\rm{tot}}^0&=\lim_{\omega\to0^+}\Re\sigma_{\rm{tot}}(\omega)\;,\label{SigmaTot0}\\
\sigma_{\rm{mag}}^0&=\lim_{\omega\to0^+}\Re\sigma_{\rm{mag}}(\omega)\;,\label{SigmaMag0}\\
\sigma_{\rm{\rm{chi}}}^0&=\lim_{\omega\to0^+}\Re\sigma_{xy}(\omega)\;.\label{SigmaChiral0}
\end{align}
Let us finally note that due to the local response with $\q=0$, all intraband contributions are contained in the Drude term, and the interband term is often referred to as the regular contribution. In the Kramers-Kronig relation, though, only the interband (regular) term enters due to $\Im\chi_{\alpha\beta}(\omega=0)=0$.

\begin{figure*}[t]
\includegraphics[width=0.68\columnwidth]{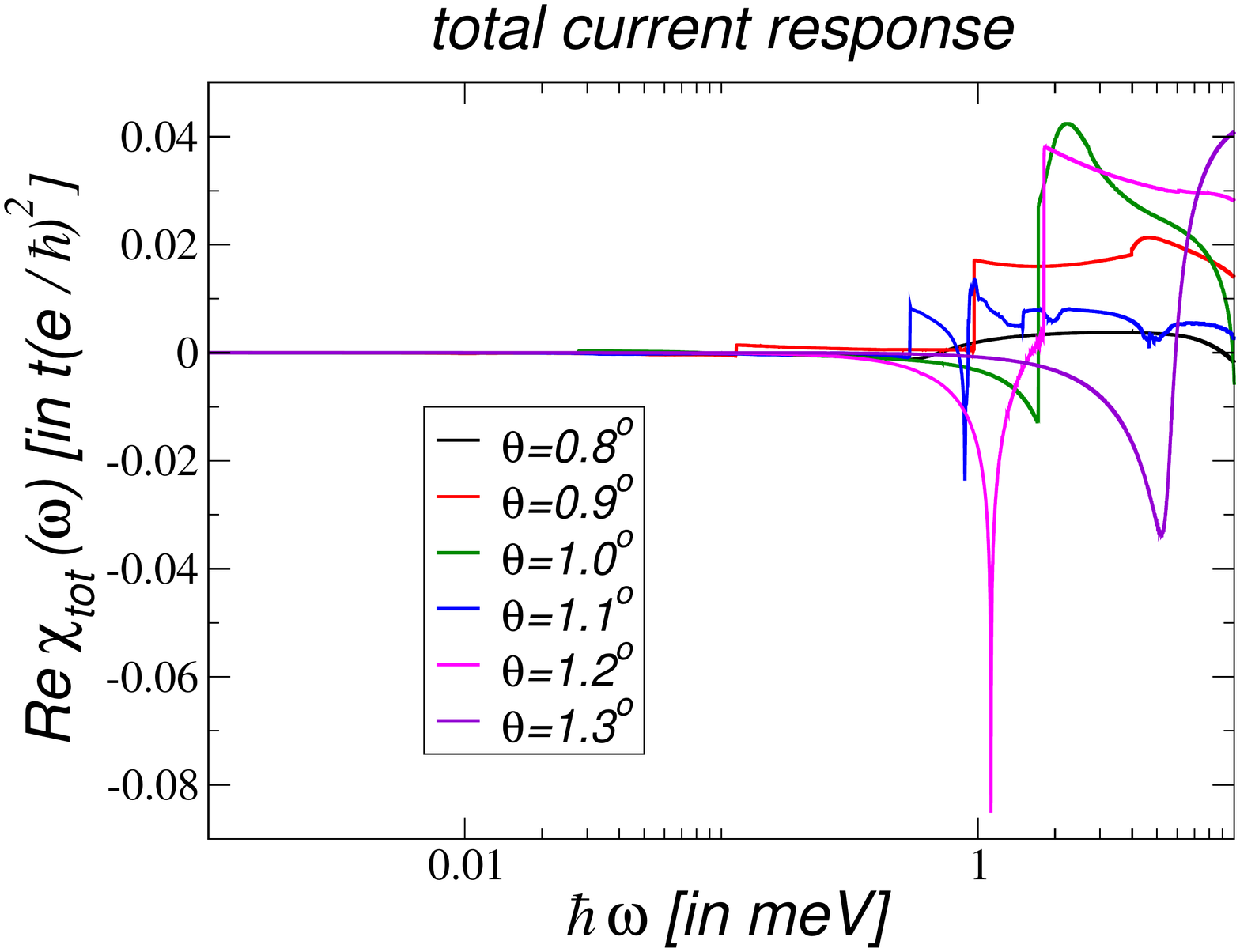}
\includegraphics[width=0.68\columnwidth]{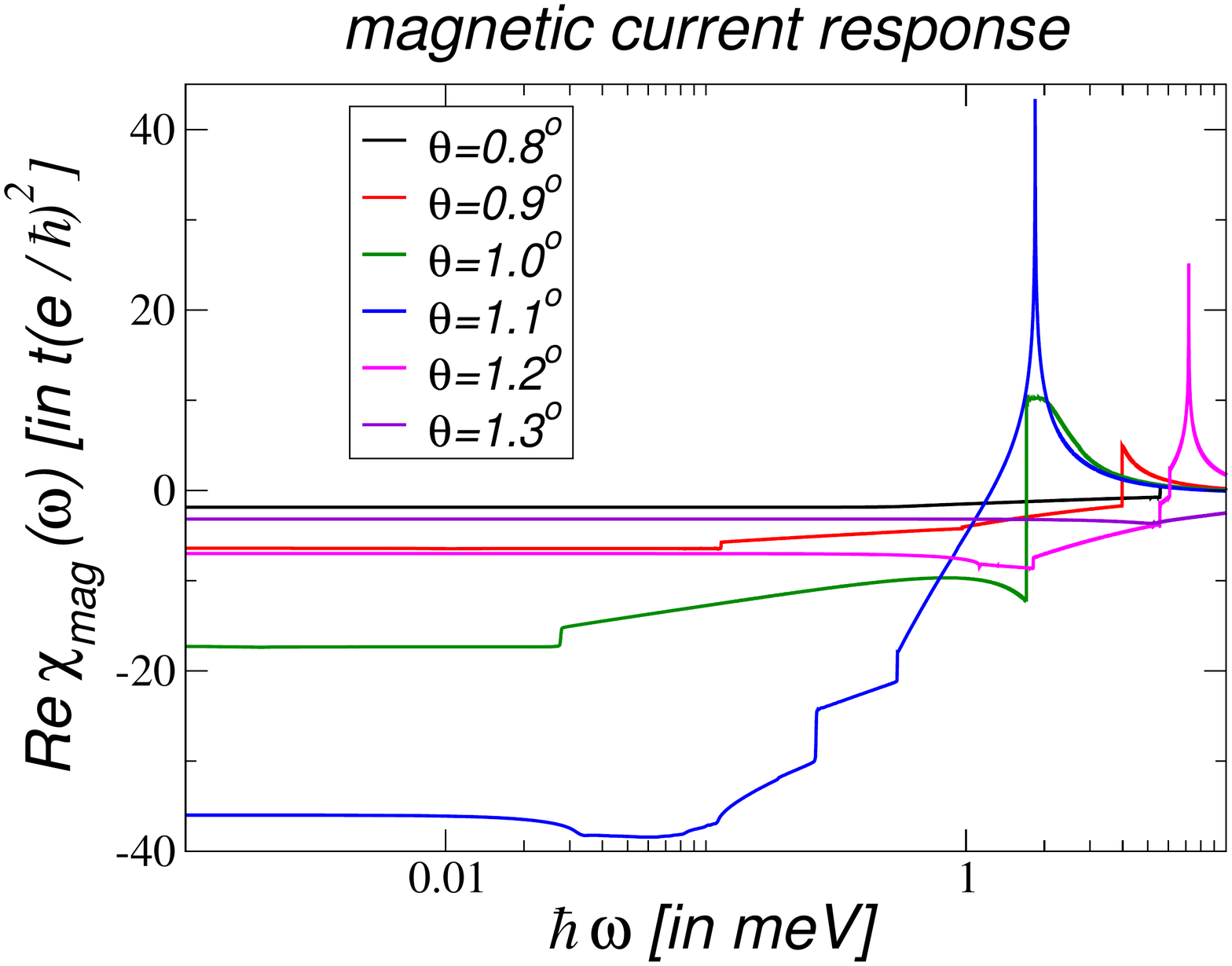}
\includegraphics[width=0.68\columnwidth]{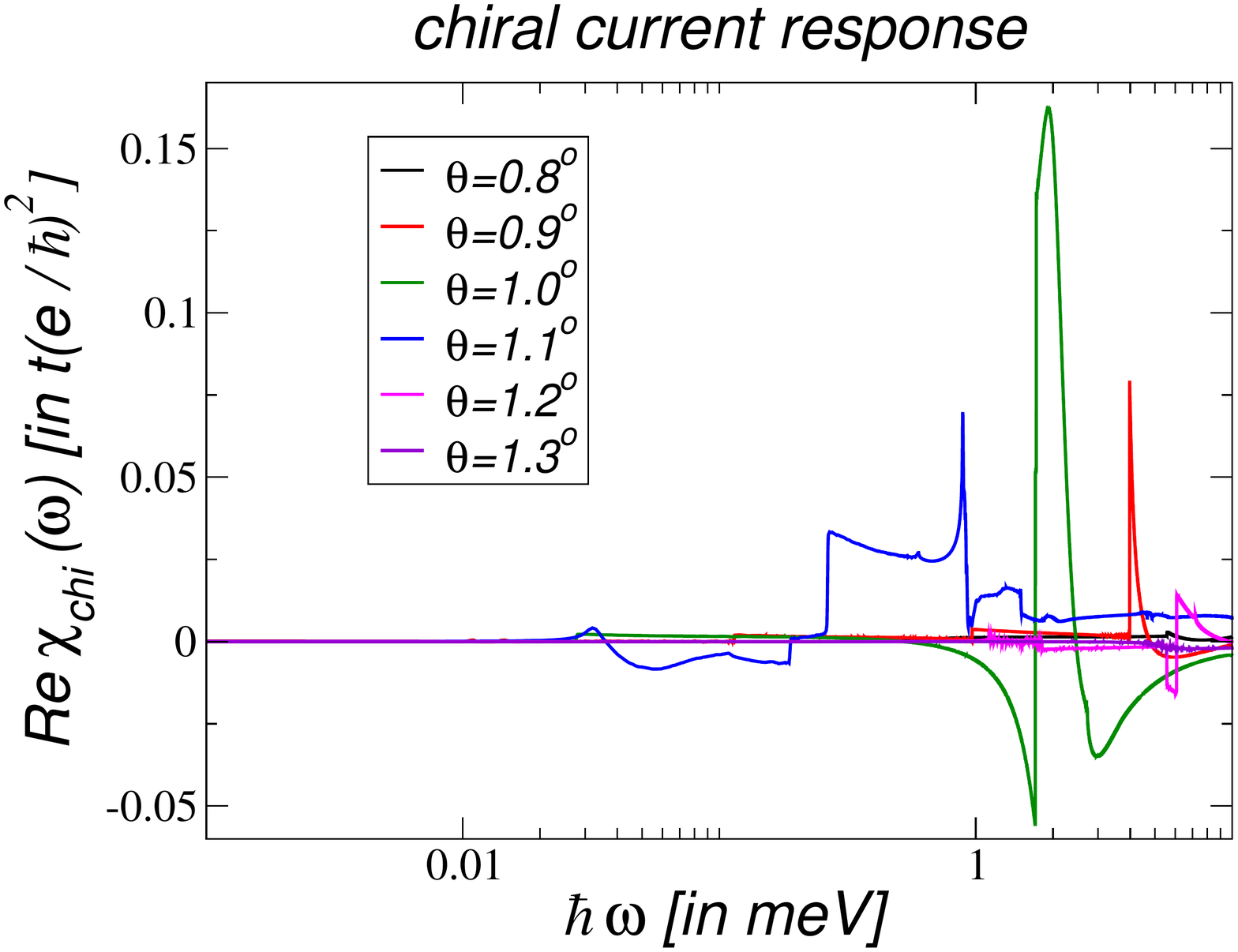}
\caption{\label{FullResponseRe} Reactive response of the total (left), magnetic (center), and chiral (right) current for the symmetric continuum model with $\kappa=1$ in Eq. (\ref{Hopping}) at various twist angles around the magic angle $\theta_m\simeq1.03^\circ$. The real current susceptibilities Re$\chi_{\rm{tot}}$, Re$\chi_{\rm{mag}}$, and Re$\chi_{\rm{chi}}$ are given in units of $t\frac{e^2}{\hbar^2}$ with $t=2.78$eV.}
\end{figure*}

\subsection{Dissipative response}
In Fig. \ref{FullResponseIm}, the dissipative response of the electric, magnetic and chiral currents at the neutrality point is shown in terms of the real part of the conductivity for angles around the magic angle $\theta\simeq1.03^\circ$. It is numerically obtained from Eq. (\ref{RealConductivity}) following the recipe outlined in the Supplementary Information, see also Ref. \onlinecite{Wiesenekker88}.  

The total optical response for $\omega\to0$ is characterized for all twist angles by the universal conductivity of two uncoupled graphene layers, $2\sigma_G$, with $\sigma_G=\frac{g_sg_v}{16}\frac{e^2}{\hbar}$; see inset of the left panel of Fig. \ref{FullResponseIm}. This is an important consistency check as in the low-frequency regime, the conductivity is entirely determined by the Dirac cone.\cite{Nair08,Kuzmenko08,Falkovsky07,Stauber08} The $1/\omega$-prefactor of the sums on the right-hand side of Eq. (\ref{RealConductivity}) is thus compensated by the weight of the corresponding Fermi line whose circumference is also proportional to $\omega$. In addition, two other universal plateaus emerge at larger frequencies; see the left panel of Fig. \ref{FullResponseIm}. These features will be discussed in  Sec. \ref{SecScaling}.

In the center panel of Fig.  \ref{FullResponseIm}, the dissipative magnetic conductivity is shown. As in the case of the total response, there are plateau for $\omega\to0$ that strongly increase around the magic angle, reaching values larger than $10^6\sigma_G$. This might also be the origin of the large orbital $g$-factor seen experimentally in twisted bilayer graphene.\cite{Li20,Sharpe21,Tschirhart21}

In the right panel of Fig.  \ref{FullResponseIm}, the dissipative chiral conductivity is shown. Again, there are plateaus marked by the Dirac regime whose values change sign at $\theta\simeq1.08^\circ$. Interestingly, this is the angle where the spectrum displays an approximate $C_6$-symmetry at each valley which renders the chiral Drude weight zero even for relatively large finite chemical potential $|\mu|\lesssim75$meV as discussed in Ref. \onlinecite{Stauber20C}.  

\subsection{Reactive response}
In Fig. \ref{FullResponseRe}, the reactive response of the total, magnetic and chiral current at the neutrality point is shown for angles around the magic angle. It is obtained from Eq. (\ref{DissipativeResponse}) via the Kramers-Kronig relation of Eq. (\ref{KramersKronig}). For this, the dissipative part needs to be determined up to a frequency $\omega_\Lambda$ for which $\Re\sigma_{\rm{tot}}(\omega\gtrsim\omega_\Lambda)\approx2\sigma_G$, $\Re\sigma_{\rm{mag}}(\omega\gtrsim\omega_\Lambda)\approx2\sigma_G$, $\Re\sigma_{\rm{chi}}(\omega\gtrsim\omega_\Lambda)\approx0$.\cite{Stauber13} These high-frequency limits represent the response of two uncoupled layers and are also a consequence of the optical sum-rule.\cite{Sabio08} 

In the left panel of Fig. \ref{FullResponseRe}, the real part of the total current response is shown. It must be zero for $\omega\to0$ as there is no excess charge in the system,\cite{Stauber20C} and we can adjust small numerical errors.\footnote{Due to our numerical procedure, there is some uncertainty in defining the cutoff-frequency and values between $D_{\rm{tot}}=-0.001t\frac{e^2}{\hbar^2}$ ($\theta=1.3^\circ$) and $D_{\rm{tot}}=0.004t\frac{e^2}{\hbar^2}$ ($\theta=1.0^\circ$) are obtained.} These shifts are also introduced to $\chi_{\rm{mag}}$ even though this does hardly have an effect as the absolute values are much higher.

In the center panel of Fig. \ref{FullResponseRe}, the real part of the magnetic current response is shown. We note that there is a non-monotonic behavior with respect to the twist angle, i.e., even though the dissipative magnetic response is peaked around the magic angle $\theta_m\simeq1.03^\circ$, see center panel of Fig. \ref{FullResponseIm}, the reactive response is not peaked at magic angle, but reaches a maximum around $\theta\simeq1.1^\circ$. This is due to the fact that for these angles, the magnetic response reaches very high values at finite frequencies with $0.1\leq\omega\leq1$meV that yield the large response due to the integration of Eq. (\ref{KramersKronig}). In Sec. \ref{SecScaling}, though, we will argue that there is a finite domain of twist angles in the immediate vicinity of the magic angle for which the magnetic current response becomes maximal and even diverges.

In the right panel of Fig. \ref{FullResponseRe}, the real part of the chiral current response is shown. It must be zero for $\omega\to0$ as there is no excess charge in the system,\cite{Stauber20C} and we can adjust small numerical errors.\footnote{Due to our numerical procedure, there is some uncertainty in defining the cutoff-frequency and values between $D_{\rm{chi}}=-0.0025t\frac{e^2}{\hbar^2}$ ($\theta=1.1^\circ$) and $D_{\rm{chi}}=0.0025t\frac{e^2}{\hbar^2}$ ($\theta=1.0^\circ$) are obtained.} For $\theta=1.0^\circ$, the maximal values can be as large as $\chi_{\rm{chi}}=0.15t\frac{e^2}{\hbar^2}$ at $\hbar\omega\cong1.9\mu$eV.
\subsection{Discussion on the Condon instability} 
\label{CondonI}
There has been considerable interest in finding systems with a symmetry-broken ground-state due to photon-condensation, the so-called Condon instability.\cite{Andolina20,Nataf19,Guerci20} In bilayer systems, this instability can also be discussed by calculating the magnetic response $D_{\rm{mag}}$. Within the random-phase approximation, the response must reach a critical value $D_{\rm{mag}}^C$ with
\begin{align}
\label{Condon}
\frac{\mu_0a}{4}D_{\rm{mag}}^C=-1\;,
\end{align}
where $\mu_0$ denotes the magnetic permeability.\cite{Sanchez21}  

For AA-stacked graphene, this limit is reached due to the logarithmic divergence of the magnetic susceptibility.\cite{Sanchez21} However, the response of twisted bilayer graphene is generally too weak to reach the instability, i.e., including damping, one obtains $D_{\rm{mag}}=-6.6t\frac{e^2}{\hbar^2}$.\cite{Stauber18,Stauber18b} Our refined calculations without damping now yield a significantly lower bound for $\theta=1.1^\circ$ with $D_{\rm{mag}}=-36t\frac{e^2}{\hbar^2}$. In the above units, this translates to $D_{\rm{mag}}\approx0.008(\mu_0a)^{-1}$ and we have $D_{\rm{mag}}/D_{\rm{mag}}^C\approx0.002$. This is still far away from a possible Condon transition. However, in Sec. \ref{SecScaling}, we will find a Condon instability in the immediate vicinity of the magic angle by employing a scaling approach.

\begin{figure*}[t]
\includegraphics[width=0.68\columnwidth]{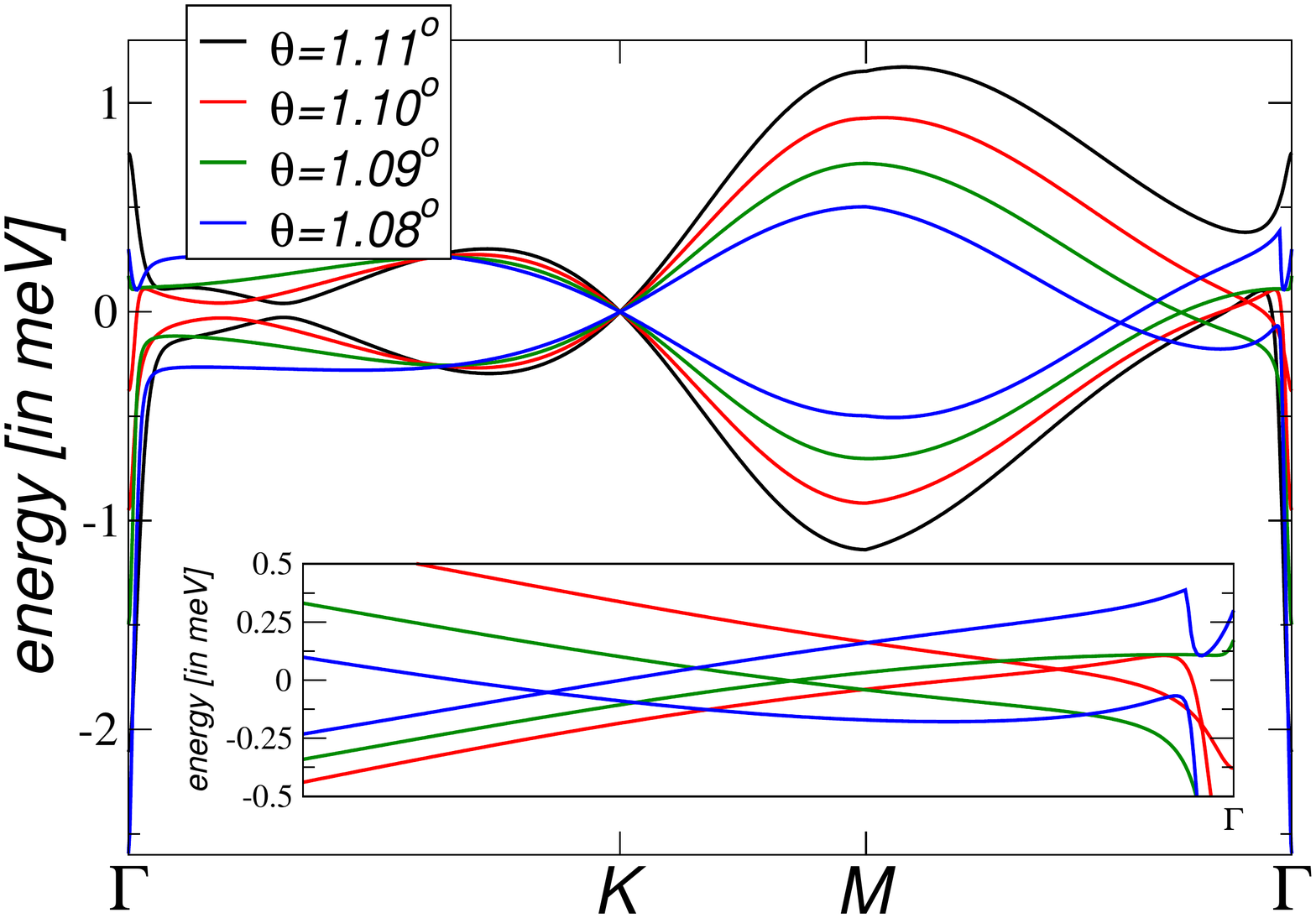}
\includegraphics[width=0.68\columnwidth]{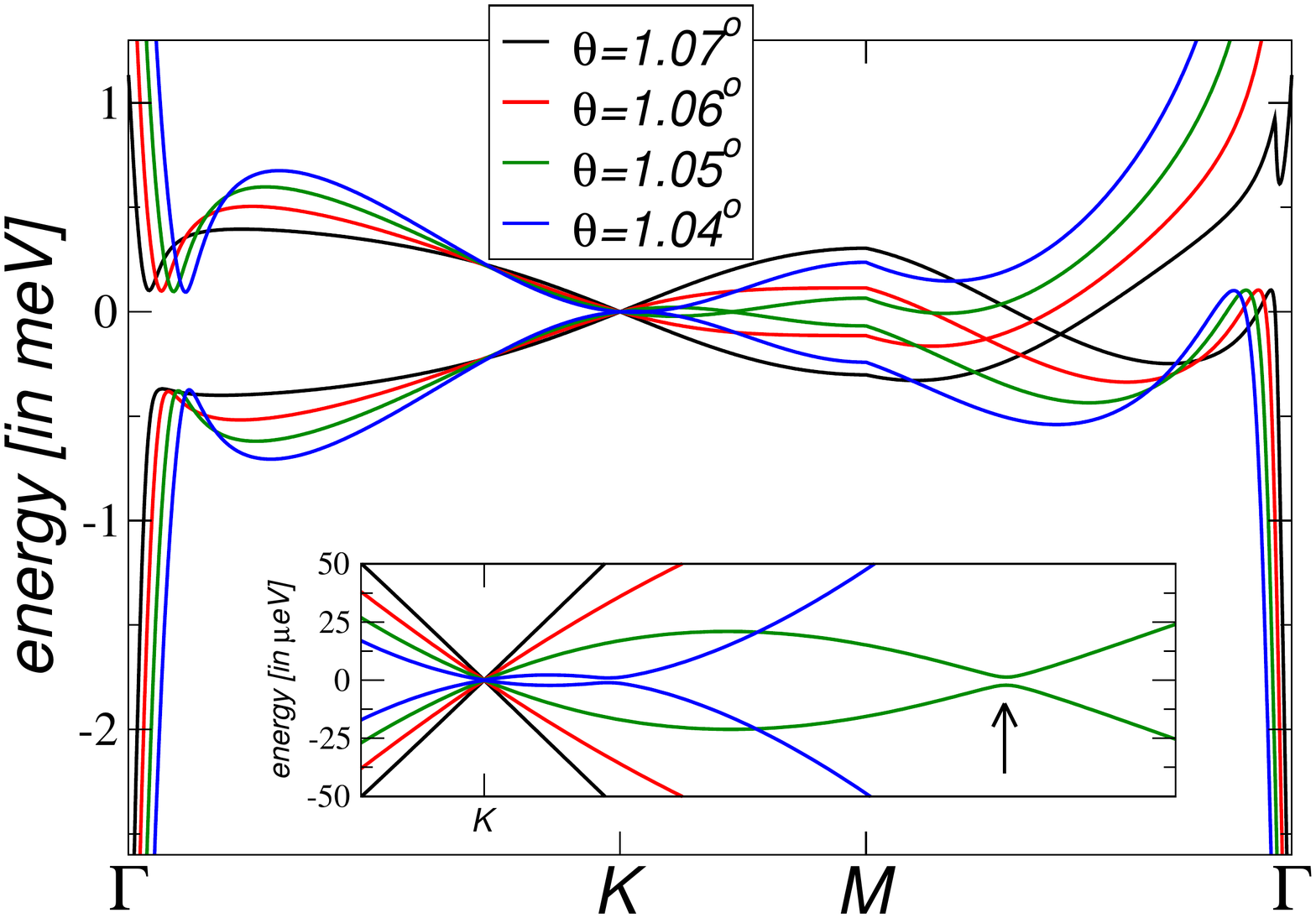}
\includegraphics[width=0.68\columnwidth]{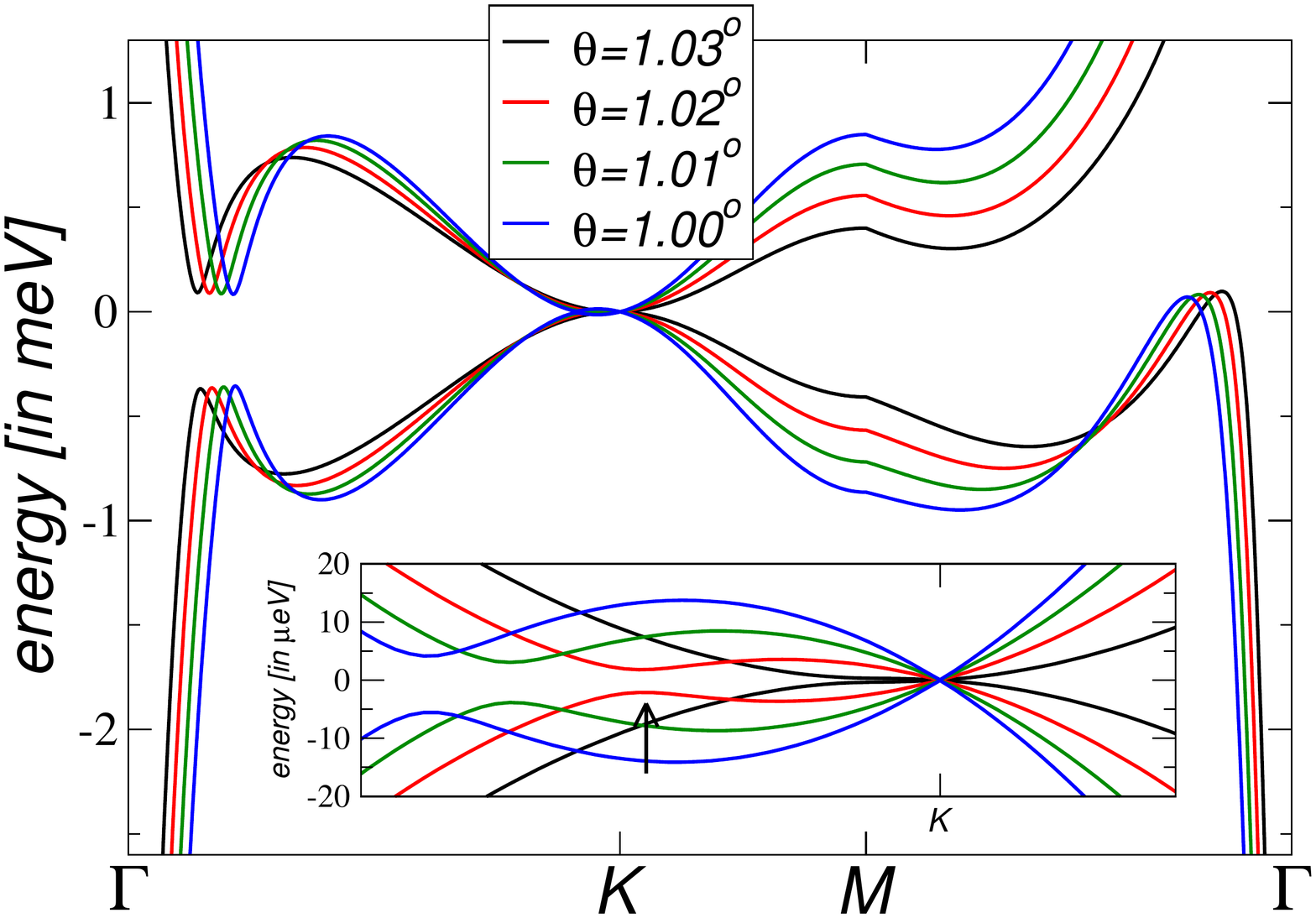}
\caption{\label{bands} Band-structure of the two flat bands around charge neutrality of the continuum model with $\kappa=1$ in Eq. (\ref{Hopping}) for various twist angles around the magic angle $\theta_m\sim1.032^\circ$. In the left panel, the band-structure with smallest band-width is shown. In the center panel, one can observe the avoided crossings along the $KM$-direction, whereas in the right panel the avoided crossings are along the $\Gamma K$-direction; see insets. The arrows indicate the avoided crossings for $\theta=1.05^\circ$ and $\theta=1.02^\circ$.}
\end{figure*}

We can compare our results also with previously reported values for the magnetic susceptibility.\cite{Guerci20} The static magnetic susceptibility $\chi_{\rm{mag}}^0$ is directly related to the magnetic Drude weight at the neutrality point and given by $\chi_{\rm{mag}}^0=\frac{a^2}{4}D_{\rm{mag}}$.\cite{Stauber18b} With $D_{\rm{mag}}=-6.6t\frac{e^2}{\hbar^2}$, this yields $\chi_{\rm{mag}}^0=0.02\frac{\mu_B}{\rm{nm}^2T}$ with $\mu_B$ the Bohr magneton.  This value, obtained for $\kappa=1$, is slightly larger than the one reported in Ref. \onlinecite{Guerci21} for the continuum model with $\kappa=0.2$. 

With $D_{\rm{mag}}\approx-36t\frac{e^2}{\hbar^2}$, see central panel of Fig. \ref{FullResponseRe}, we obtain for the static magnetic susceptibility an even larger value of $\chi_{\rm{mag}}^0=0.12\frac{\mu_B}{\rm{nm}^2T}$. This amounts to $18\mu_B$ per moir\'e cell only due to the orbital motion of counter-propagating electrons. This purely quantum mechanical effect is remarkable as no charge excitations are involved. 
 
\section{Optical response at the magic angle}
\label{SecScaling}
In this section, we discuss the optical response in the immediate vicinity of the magic angle, i.e., we will highly zoom into this region of possible twist angles. As we will see, for any angle one can find an energy regime which is still characterized by the Dirac cone, i.e., one will never be exactly {\it at} the magic angle just as one can never approach an irrational number. Furthermore, other plateaus develop which can be anticipated from the band structure which shall be discussed before we describe the scaling relations.  

\subsection{Bands around the magic angle}
In Fig. \ref{bands}, the band structure for the symmetric model is shown for different twist angles around the magic angle $\theta_m\simeq1.032^\circ$ where the linear dispersion (Fermi velocity) at the $K$-point vanishes.\cite{SongZhida19,Hejazi19,Koshino19B} Notice that this does not coincide with the smallest band-width condition which would yield a magic twist-angle of $\theta_m^*\simeq1.11^\circ$. This can be appreciated on the left panel of Fig. \ref{bands} where a new regime starts with an accidental crossing on the $\Gamma M$-direction. 

In the center and right panel of Fig. \ref{bands}, we see the evolution towards the magic angle from above and below, respectively. Most notably, there is an avoided crossing that is moving closer to the $K$-point when approaching the magic angle which is highlighted in the insets. For even smaller angles, a stable band-inversion emerges with the avoided crossing moving outward and eventually inward again to form the second magic angle. The evolution in $\theta$ around the second magic angle at $\theta\sim0.49^\circ$, however, is qualitatively different. 

\subsection{Scaling in the immediate vicinity of the magic angle}
The universal conductivity of graphene for small frequencies, $\sigma_G=\frac{e^2}{4\hbar}$, is due to the perfect cancellation between the transition-matrix element and the Fermi velocity.\cite{Nair08,Kuzmenko08,Peres08} This is also the case for the total conductivity of twisted bilayer graphene for transitions around the Dirac cones. Considering different quantities such as the magnetic absorption related to $\Re\sigma_{\rm{mag}}$ or the chiral absorption related to $\Re\sigma_{\rm{chi}}$ will not show this cancellation and we expect the following relations for $\omega\to0$:\cite{Bistritzer11}
\begin{align}
\label{OpticalConductivities}
\sigma_{\rm{mag}}^0=\sigma_G\left(\frac{v_{\rm{mag}}}{v_F}\right)^2\;,\;\sigma_{\rm{chi}}^0=\sigma_G\left(\frac{v_{\rm{chi}}}{v_F}\right)^2
\end{align}
Above, we defined suitable velocities that characterize the magnetic and chiral excitations. 

\begin{figure*}[t]
\includegraphics[width=0.68\columnwidth]{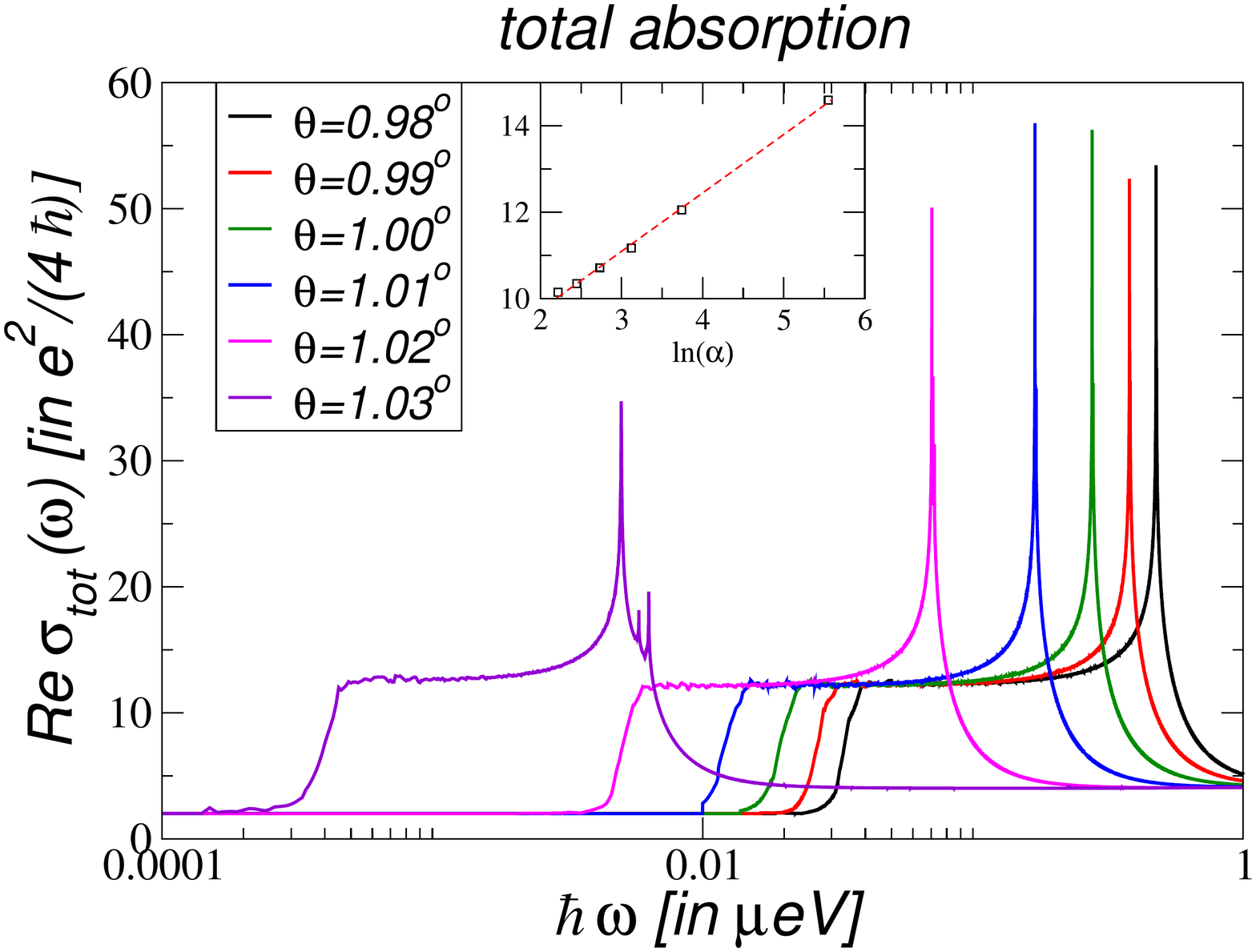}
\includegraphics[width=0.68\columnwidth]{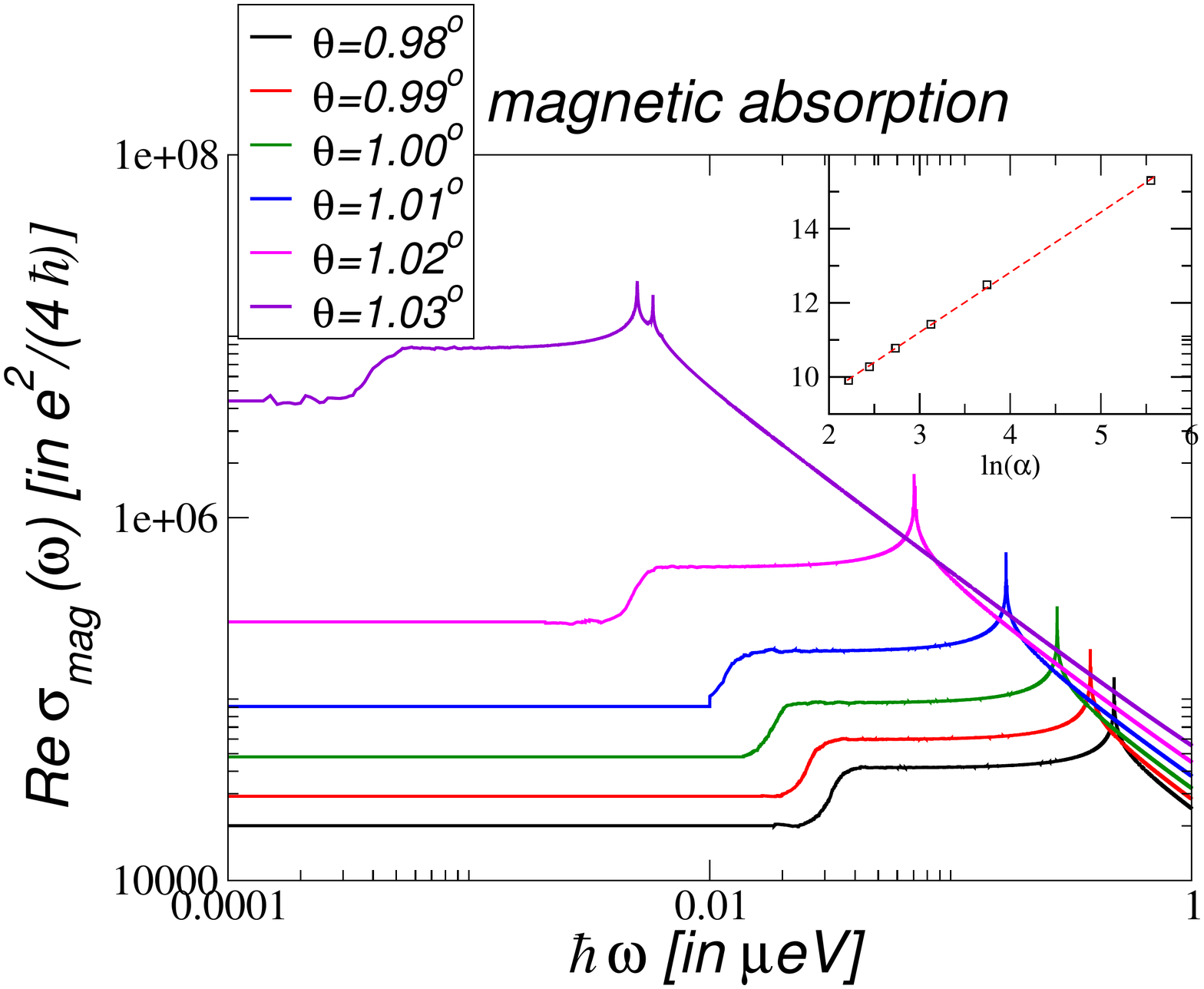}
\includegraphics[width=0.68\columnwidth]{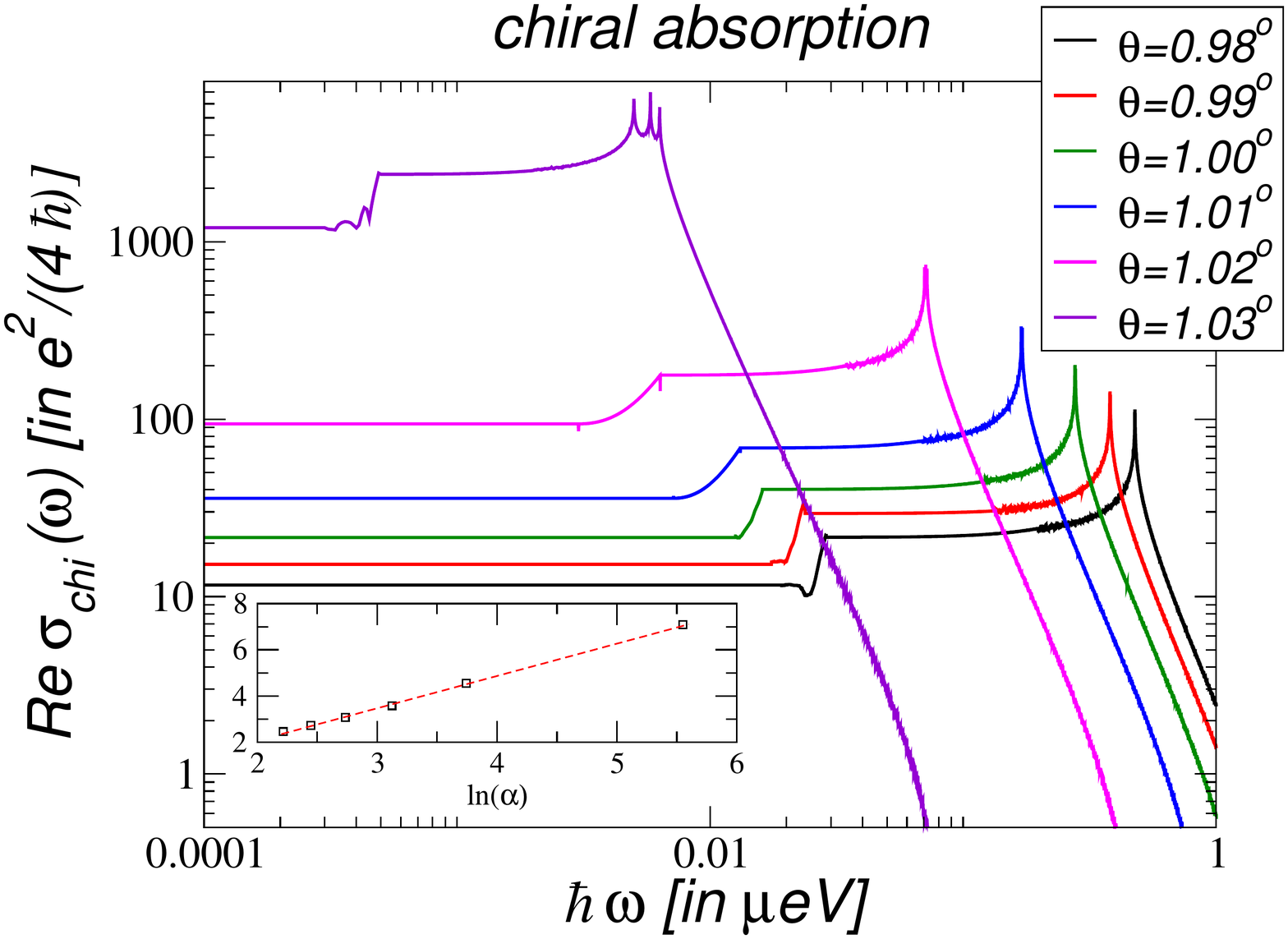}
\caption{\label{MagicClose} The optical response for twist angles below the magic angle $\theta_m=1.032^\circ$. The insets shows the logarithm of the Dirac regime (left panel) and the optical response functions $\sigma_{\rm{mag}}$ (center panel) and $\sigma_{\rm{chi}}$ (right panel) at $\omega=0$ as function of the effective parameter ln$\alpha$ defined in Eq. (\ref{alphaDef}).}
\end{figure*}

As the Fermi velocity vanishes at the magic angle, Eq. (\ref{OpticalConductivities}) suggest that the magnetic and chiral absorption diverge. Our numerical calculations confirm precisely this, as can be seen in Fig. \ref{MagicClose}, where we show the three response functions for twist angles below the magic angle $\theta_m\sim1.032^\circ$. Whereas the absorption shows universal behavior, the magnetic as well as the chiral absorption diverge. 

In order to discuss the scaling behavior of these quantities, we introduce the effective parameter 
\begin{align}
\label{alphaDef}
\alpha=\frac{\alpha_{\theta_i}-\alpha_m}{\alpha_m}\approx\frac{\theta_m-\theta_i}{\theta_i}\;, 
\end{align}
with $\alpha_{\theta_i}=\frac{\sqrt{A_i/3}}{2\pi}\frac{t_\perp}{t}$ and $\alpha_m\sim0.605$ (for $i=31.54$). 

First, we investigate the scaling of the Dirac regime $\epsilon_D$ that is defined by the abrupt increase of the absorption from $2\sigma_G$ to $12\sigma_G$. As is shown in the inset of the left panel of Fig. \ref{MagicClose}, there is a linear behavior of $\ln\epsilon_D$ as function of $\ln\alpha$, leading to $\epsilon_D\simeq0.88\alpha^{\gamma_{\epsilon}}\mu$eV with $\gamma_{\epsilon}\simeq1.35\pm0.04$. 

Along the same lines, we obtain the scaling behavior of the magnetic and chiral absorption plateau for $\hbar\omega<\epsilon_D$ as
\begin{align}
\sigma_{\rm{mag}}^0=570\sigma_G\alpha^{-\gamma_{\rm{mag}}}\;,\;\sigma_{\rm{chi}}^0=0.48\sigma_G\alpha^{-\gamma_{\rm{chi}}}\;.
\end{align}
with $\gamma_{\rm{mag}}\simeq1.62\pm0.03$ and $\gamma_{\rm{chi}}\simeq1.41\pm0.02$. 

It is generally argued that the Fermi velocity scales linearly in $\alpha$.\cite{Lopes07,Bistritzer11,Lopes12,Peres19,Watson21,Becker21} This implies that $v_{\rm{mag}}=24v_F\alpha^{1-\gamma_{\rm{mag}}/2}$. In addition, the chiral velocity must also tend to zero at the magic angle as $v_{\rm{chi}}=0.7v_F\alpha^{1-\gamma_{\rm{chi}}/2}$.

\subsection{Condon instability at the magic angle}
As mentioned in Sec. \ref{CondonI}, in AA-stacked bilayer graphene there is a Condon instability at $T\sim0$. Since in twisted bilayer graphene, the electronic wave functions at the magic angle are highly localized around  the AA-stacked islands,\cite{Trambly10} there might be the possibility of a Condon instability in twisted bilayer graphene around $\theta_m$.\cite{Guerci21} 

The imaginary part of the conductivity or magnetic Drude weight is obtained from the Kramers-Kronig relation which can be split into the following two contributions:
\begin{align}
D_{\rm{mag}}&=\frac{2}{\pi}\left[2\omega_\Lambda\sigma_G-\int_0^{\omega_\Lambda} d\omega \Re\sigma_{\rm{mag}}(\omega)\right]\label{SplittingD}\\
&=\frac{2}{\pi}\left[2\omega_\Lambda\sigma_G-\left(\int_0^{\omega_D^{\rm{mag}}}+\int_{\omega_D^{\rm{mag}}}^{\omega_\Lambda}\right) d\omega \Re\sigma_{\rm{mag}}(\omega)\right]\nonumber\\
&=D_{\rm{mag}}^*+D_{\rm{mag}}^{reg}\nonumber\;,
\end{align}
where $\omega_\Lambda$ denotes the high-frequency cutoff and $\omega_D^{\rm{mag}}>0$ is the smallest frequency after the van Hove singularity for which $\sigma_{\rm{mag}}^0=\sigma_{\rm{mag}}(\omega_D^{\rm{mag}})$, i.e., for $\theta=1.03^\circ$, this gives $\hbar\omega_D^{\rm{mag}}\approx0.01\mu$eV.

The second term $D_{\rm{mag}}^{reg}$ is assumed to be regular. The possible divergent contribution at the magic angle, $D_{\rm{mag}}^*$, can be estimated as follows: 
\begin{align}
D_{\rm{mag}}^*=-\frac{2}{\pi}\int_0^{\omega_D^{\rm{mag}}} d\omega \sigma_{\rm{mag}}(\omega)\sim-\alpha^{-\gamma_{\rm{mag}}+\gamma_{\epsilon}^{\rm{mag}}}
\end{align}
The exponent $\gamma_{\epsilon}^{\rm{mag}}$ is again obtained from a linear fit of a log-log plot and we obtain $\gamma_{\epsilon}^{\rm{mag}}\simeq1.41\pm0.04$. We thus find a divergence at the magic angle that scales like $D_{\rm{mag}}^*\sim-\alpha^{-\delta_{\rm{mag}}}$ with $\delta_{\rm{mag}}=0.21\pm0.05$. Since the Condon instability is marked by $D_{\rm{mag}}\sim D_{\rm{mag}}^C=\frac{4}{\mu_0a}$, see Eq. (\ref{Condon}),  there will be a symmetry-broken ground-state with orbital magnetic domains at the magic angle. 

The presence of an instability due to transverse current fluctuations in a {\it non-interacting} model of Eq. (\ref{Hamiltonian}) is a remarkable result and we are not aware of any other non-interacting model that exhibits a symmetry-broken ground-state other than $AA$-stacked bilayer graphene.\cite{Sanchez21} Let us finally note that the total chiral Drude weight $D_{\rm{chi}}$ has to vanish at the neutrality point due to gauge symmetry.\cite{Stauber18b,Stauber20} 
\subsection{Mapping to effective model}
The absorption spectrum in the immediate vicinity of the magic angle can approximately be understood from the universal conductivity formula\cite{Stauber15} of a general dispersion $\epsilon_\k\propto|\k|^\nu$
\begin{align}
\sigma(\omega)=\frac{g_sg_vg_\ell g_{C_3}}{16}\nu\frac{e^2}{\hbar}=g_\ell g_{C_3}\sigma_G\;.
\end{align}
Above, we introduced the usual spin, valley, and layer-degree of freedom, but also a possible $g_{C_3}$ degeneracy which takes the value 3 in case of an explicit 3-fold degeneracy (otherwise it is 1). In the following, we will discuss the results in units of the universal conductivity of graphene $\sigma_G=\frac{g_sg_v}{16}\frac{e^2}{\hbar}$. Notice that we introduce here explicitly the degeneracy factors which are usually set to $g_s=g_v=2$.

At low frequencies, there will in principle always be a regime where the absorption is governed by the universal absorption of Dirac Fermions with $\nu=1$ and we have $\sigma(\omega)=2\sigma_G$. For twist angles in the immediate vicinity of the magic angle, the plateau of a single quadratic dispersion relation with $\nu=2$ is obtained with $\sigma(\omega)=4\sigma_G$, seen in the left panel of Fig. \ref{MagicClose} for $\theta=1.03^\circ$ for $0.01\mu$eV$\lesssim\epsilon\lesssim1\mu$eV.

Between these plateaus, a new plateau emerges with $\sigma_0=12\sigma_G$, because a new absorption channel opens at the frequency of the avoided crossing as seen in the inset of the center and right panels of Fig. \ref{bands} and marked by arrows for $\theta=1.05^\circ$ and $\theta=1.02^\circ$. Even though the band minima are elongated, as a first approximation they can be assumed to be a quadratic dispersion and due to the $C_3$-symmetry, there are three of them for each Dirac point. We thus numerically obtain $\sigma(\omega)=12\sigma_G$.\footnote{This plateau $12\sigma_G$ is only obtained for twist angles which are already sufficiently close to the magic angle, i.e., the band structure for $\theta=1.05^\circ$ shows an avoided crossing, but does not reach this plateau, yet.} 
 
However, we have been neglecting the contribution of the central Dirac cone and the above qualitative discussion can be made quantitative by considering the following two-band model which was first introduced in Refs. \onlinecite{Bena11,Montambaux12,Hejazi19} for $M=0$:
\begin{align}
\label{FrenchModel}
H_\k=\frac{1}{2m}\begin{bmatrix} M & \varpi^2+\eta \varpi^*\\{\varpi^*}^2+\eta\varpi&-M\end{bmatrix}
\end{align}
where $\varpi=\hbar(k_x-ik_y)$. The model has eigenenergies $2m\epsilon_\k=\pm\sqrt{M^2+k^4+2k^3\cos(3\theta)\eta+k^2\eta^2}$ displaying trigonal warping and zeros at $|\k|=\eta$. For $M=0$, there are three nodal points which lie in the directions $\theta=\frac{2\pi n}{3}$ ($\eta<0$) and $\theta=\pi-\frac{2\pi n}{3}$ ($\eta>0$) with $n\in\mathbb{N}$. This transition can be also seen in the center and right panel of Fig. \ref{bands}, where the avoided crossing changes from the $KM$-direction (right from the $K$-point) to the $\Gamma M$-direction (left from the $K$-point), related by a $60^\circ$-rotation.
 
As shown in the Supplementary Information, the above model with $M=0$ yields $\sigma=12\sigma_G$ for small frequencies and $\sigma=4\sigma_G$ for large frequencies. The reason for not obtaining the Dirac regime $\sigma=2\sigma_G$ is because the model of Eq. (\ref{FrenchModel}) with $M=0$ does not exhibit a gap at the three nodal points with $|\k|=\eta$. 

This can partially be remedied by introducing a $\k$-dependent mass term with $M=|\hbar\k|^2$ such that the gap or Dirac-regime energy is given by $\epsilon_D=\frac{\hbar^2}{m}\eta^2$. From the numerical approach we obtain $\epsilon_D=0.88\alpha^{\gamma_\epsilon}\mu$eV. At the magic angle $\theta_m\sim1.03^\circ$, we can further extract the mass-term since $\eta=0$. Remarkably, we get $m\approx m_0$ where $m_0$ is the mass of free electrons. This allows us to connect $\eta$ to $\alpha$ of Eq. (\ref{alphaDef}):
\begin{align}
\label{Mapping}
\hbar\eta=\sqrt{0.88m\mu\rm{eV}}\alpha^{\gamma_\epsilon/2}
\end{align}
Notice that with the discussion of the dimensionless energy scale $\tilde\omega$ defined in the Supplementary Information, we would obtain the same scaling relation. Eq. (\ref{Mapping}) together with $m\approx m_0$ provides a direct mapping between the continuum model of twisted bilayer graphene and the model of Eq. (\ref{FrenchModel}) in the immediate vicinity of the magic angle in the flat-band regime.
\begin{figure*}[t]
\includegraphics[width=0.68\columnwidth]{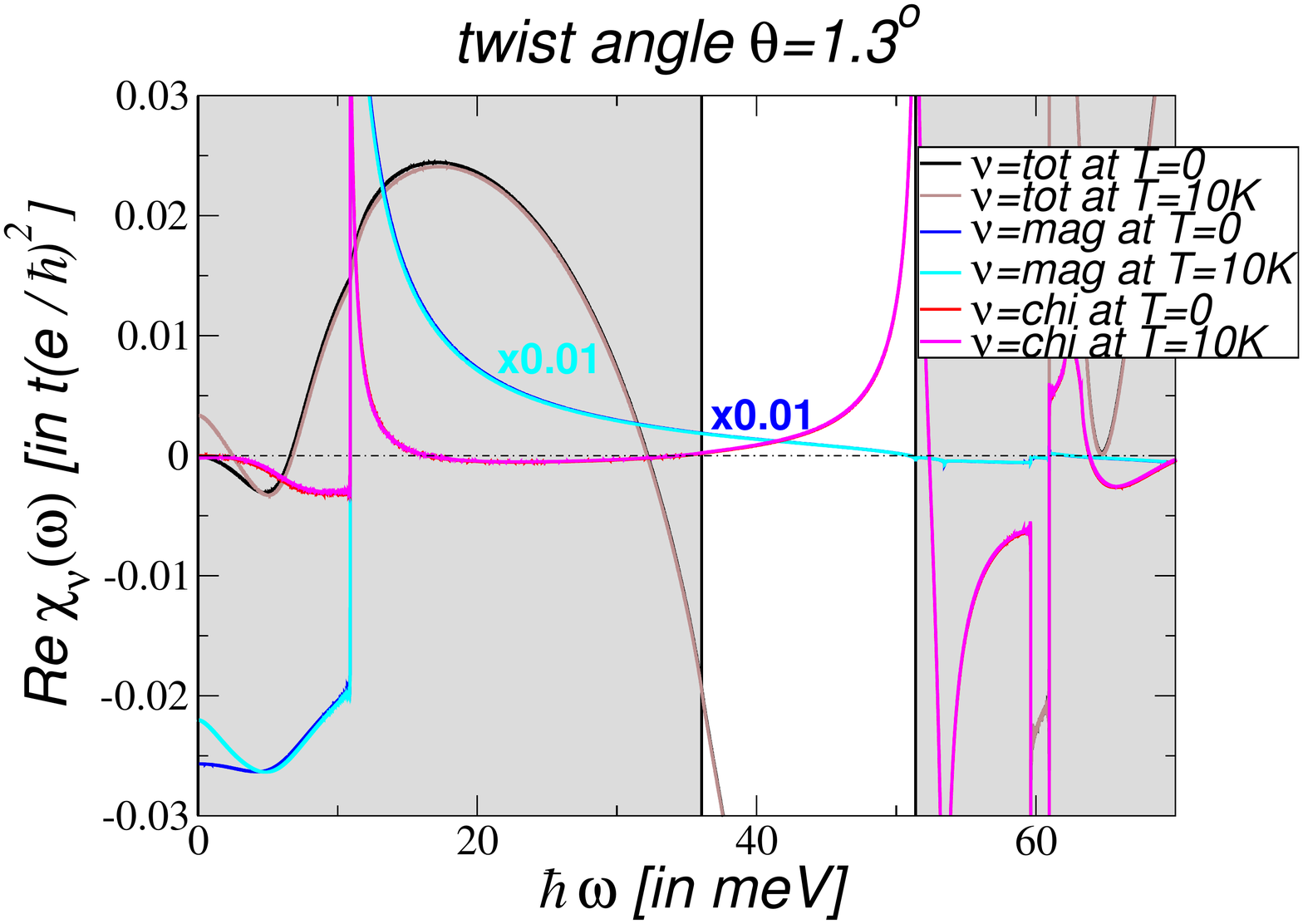}
\includegraphics[width=0.68\columnwidth]{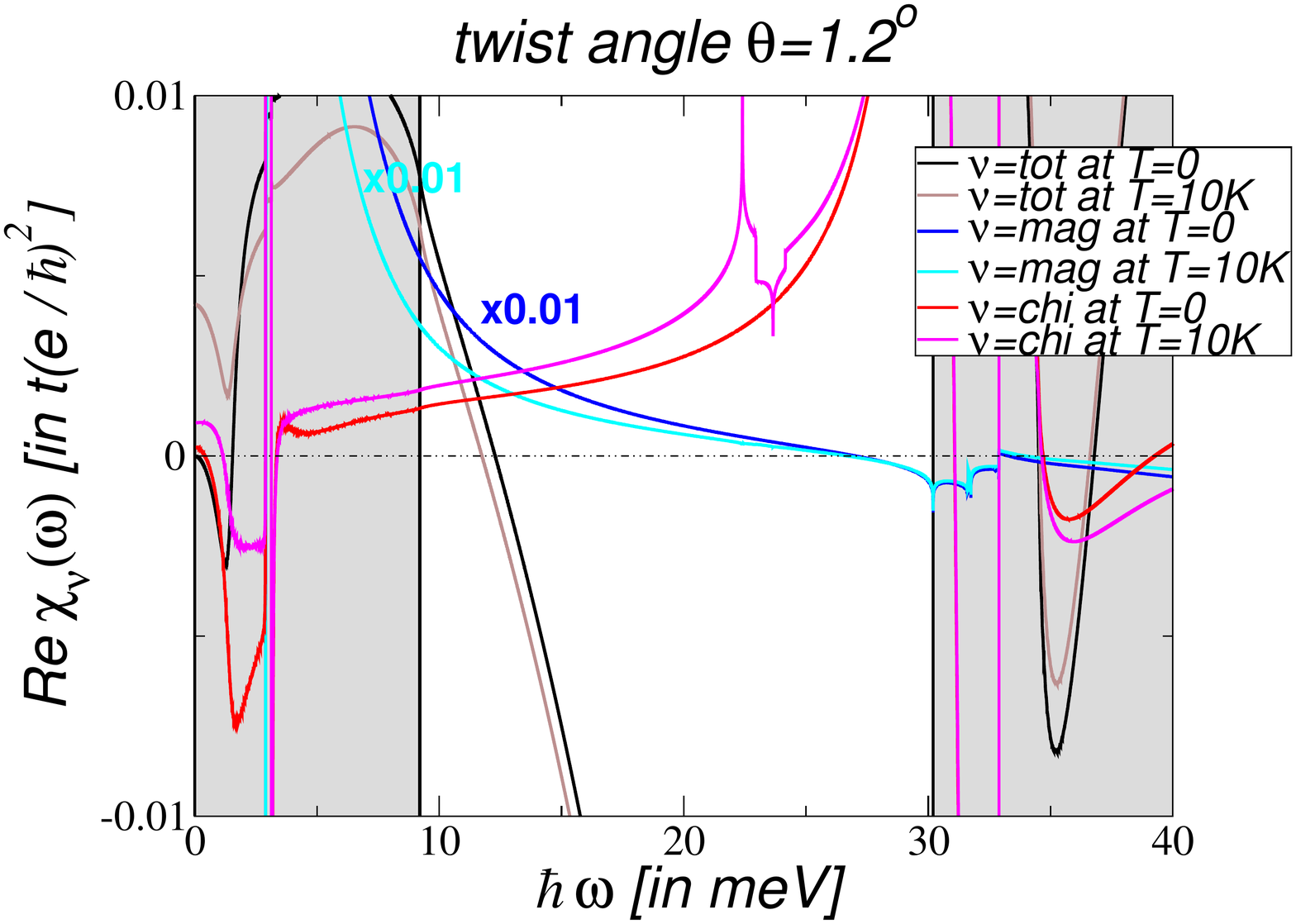}
\includegraphics[width=0.68\columnwidth]{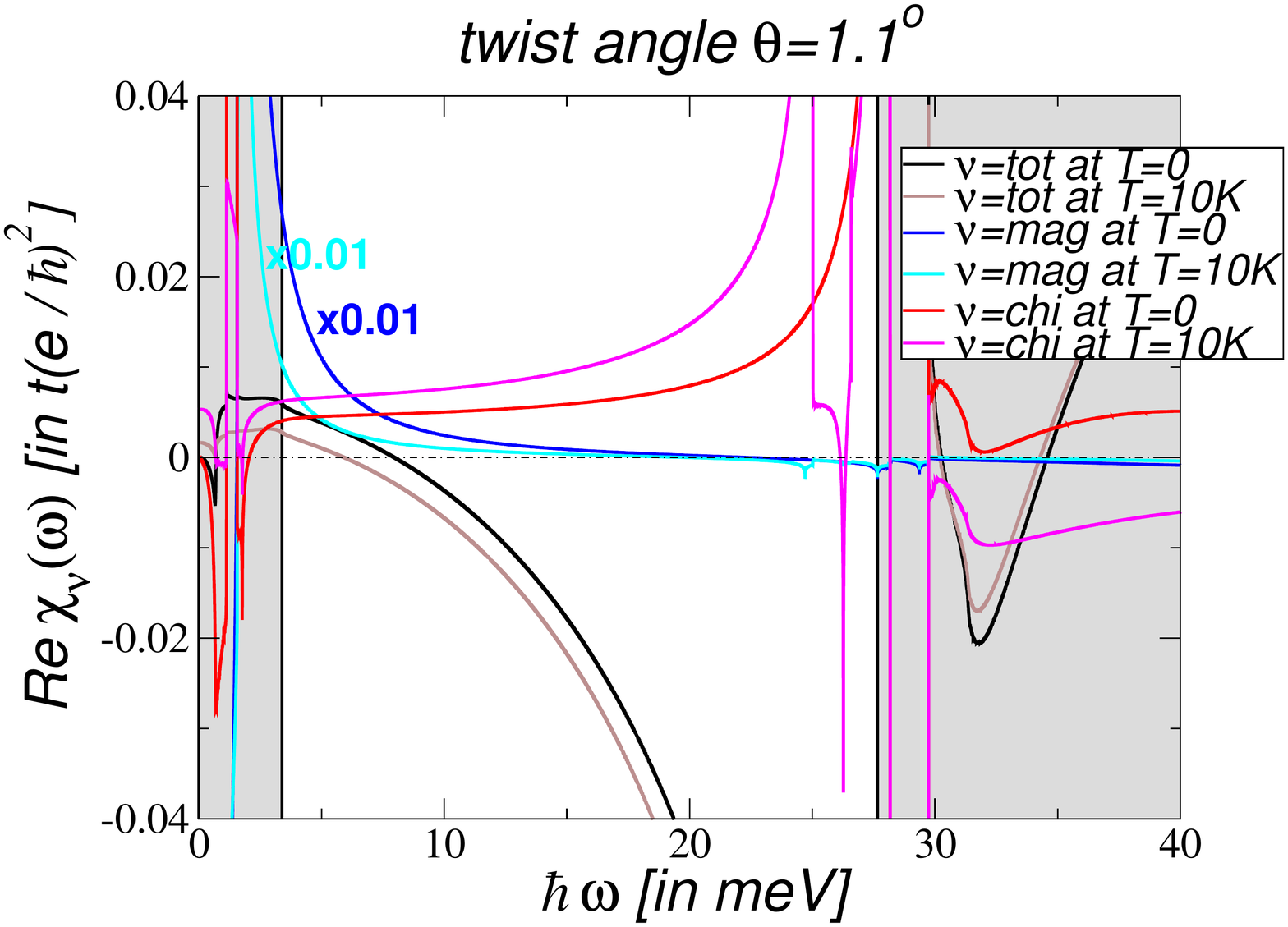}
\caption{\label{Plasmon} The real part of the current susceptibility Re$\chi_\nu(\omega)$ with $\nu=\rm{tot},\rm{mag},\rm{chi}$ of the asymmetric continuum model with $\kappa=0.8$ in Eq. (\ref{Hamiltonian}) at the neutrality point in units of $t\frac{e^2}{\hbar^2}$ for temperatures $T=0,10$K. The optical gap is indicated by the white area. Left panel: twist angle $\theta=1.3^\circ$. Center panel: twist angle $\theta=1.2^\circ$. Right panel: twist angle $\theta=1.1^\circ$.}
\end{figure*}
\section{Flat-band plasmonics}
Since twisted bilayer graphene consists of two layers, there will be two plasmonic modes. For layers far away, these modes hardly hybridize, but for an interlayer distance $a=3.5\, \mathring{\text{A}}$, anti-bonding and bonding modes emerge. Due to the long-ranged Coulomb interaction, the dispersions show square-root and linear behavior in the momentum $q$ and define the so-called optical (charge even) and acoustic (charge odd) branches, respectively. In the local approximation, they are generally given by 
\begin{align}
\label{PlasmonDispersionOp}
\omega_+^2&=\frac{\chi_{\rm{tot}}(\omega_+)q}{2\epsilon_0\epsilon}\;,\\
\label{PlasmonDispersionAc}
\omega_-^2&=\frac{\chi_{\rm{mag}}(\omega_-)aq^2}{2\epsilon_0}\;,
\end{align}
which define self-consistent equations for the plasmonic frequencies $\omega_+$ and $\omega_-$ with momentum $q$, respectively. Note that the optical mode depends on the dielectric environment through $\epsilon=(\epsilon_{up}+\epsilon_{down})/2$, but the acoustic mode does not.\cite{Stauber14} 

The plasmon dispersion does not depend on the chiral Drude weight since the non-retarded approximation does not allow for a coupling of longitudinal and transverse modes.\cite{Stauber18,Lin20} Nevertheless, the optical (acoustic) mode, usually defined by electric (magnetic) dipole oscillations, is now accompanied by parallel magnetic (electric) dipole oscillations. With the magnetic dipole related to the magnetic current as  $2\m=a\j_{\rm{mag}}\times\e_z$, this is expressed by the following relations:
\begin{align}
\e_\q\cdot\m&=-aX_{\rm{tot}}\e_\q\cdot\j_{\rm{tot}}\;,\\
4a\e_{\q_\perp}\cdot\j_{\rm{tot}}&=-X_{\rm{mag}}\e_{\q_\perp}\cdot\m\;,
\end{align}
with 
\begin{align}
X_{\rm{tot}}=\frac{\chi_{\rm{chi}}}{\chi_{\rm{tot}}}\;,\; X_{\rm{mag}}=\frac{\chi_{\rm{chi}}}{\chi_{\rm{mag}}}\;. 
\end{align}
The total current is related to the electric dipole, $\j_{\rm{tot}}=-i\omega\p$, and we have $\p\parallel\m\parallel\q$ for the optical mode and $\p\parallel\m\perp\q$ for the acoustic mode. 

The above relations are obtained from the transport equations of Eq. (\ref{CurrentSymmetries}) and hold also in the static limit, i.e., the total Drude weight $D_{\rm{tot}}$ and chiral Drude weight $D_{\rm{chi}}$ are Fermi-line properties as discussed in Ref. \onlinecite{Stauber20C}. Similar conclusions have been drawn in Refs. \onlinecite{HeWenYo20,Antebi22}.

Let us finally note, that Eqs. (\ref{PlasmonDispersionOp}) and (\ref{PlasmonDispersionAc}) can be generalized to a non-local approximation by the replacements $\chi_{\rm{tot}}(\omega)\to\chi_{\rm{tot}}(\omega,\q)$ and $\chi_{\rm{mag}}(\omega)\to\chi_{\rm{mag}}(\omega,\q)$ that leads to flat plasmonic bands.\cite{Stauber16} 
\subsection{Poynting vector}
Even though the optical and acoustic plasmon dispersions only depend on $\chi_{\rm{tot}}$ and $\chi_{\rm{mag}}$, respectively, the Poynting vector depends also on the chiral response $\chi_{\rm{chi}}$. To show this, let both modes be induced by the sheet current $j_\parallel$ parallel to the plasmon momentum $\q$, i.e., decomposing the Fourier components of the current into longitudinal and transverse parts, we have $\j=j_\parallel\e_\q+j_\perp\e_{\q_\perp}$ for layer $\ell=1$. 

For the optical mode, the sheet currents of the two layers are parallel and for the acoustic mode, the sheet currents of the two layers are anti-parallel. In the instantaneous approximation, the self-fields are purely longitudinal, and we have $\q_\perp\cdot\E^\ell=0$ as well as $\q\cdot\E^1=\q\cdot\E^2$ for the optical mode and $\q\cdot\E^1=-\q\cdot\E^2$ for the acoustic mode. This yields the relation between the longitudinal and transverse current as $j_\perp/j_\parallel=-2X_{\rm{tot}}$ and $j_\perp/j_\parallel=2X_{\rm{mag}}$ for the two modes, respectively.

We then get in the non-retarded limit, close to the sheet and up to second order in $qa$, the following expressions for the Poynting vectors of the optical (tot) and acoustic (mag) mode (see also Ref. \onlinecite{Stauber20}):
\begin{align}
\mathcal{P}_{\rm{tot}}&=\mathcal{P}_0\left(\begin{matrix}1+\gamma_{\rm{tot}}(qa)^2\\2sgn(z)X_{\rm{tot}} qa\\0\end{matrix}\right)\;,\\\mathcal{P}_{\rm{mag}}&=\mathcal{P}_0\left(\begin{matrix}4X_{\rm{mag}}^2\frac{k_0^2}{q^2}+\gamma_{\rm{mag}}(qa)^2\\-2sgn(z)X_{\rm{mag}} qa\\0\end{matrix}\right)\;,\label{Pacc}
\end{align}
with $\mathcal{P}_0=\frac{qj_\parallel^2}{2\epsilon_0\omega}$ and $\gamma_{\nu}=[1+(4X_\nu^2-1)\frac{k_0^2}{q^2}]/4$, where $k_0=\omega/c$ is the wavelength of light in free space and $\nu=\rm{tot},\rm{mag}$. This shows that the chirality modifies the plasmonic energy flux. Also note that the Poynting vector of the acoustic mode to lowest order in $aq$ and in the non-chiral limit $X_{\rm{mag}}=0$ becomes zero since this mode consists of perfectly cancelling counterpropagating current densities.

Let us now discuss the limiting case $X_\nu\ll1$ for $k_0/q\to0$ and $aq\to0$. We then have for the Poynting vectors of the optical (tot) and acoustic (mag) mode the following expressions:
\begin{align}
\mathcal{P}_{\rm{tot}}&=\mathcal{P}_0\left(\begin{matrix}1\\2sgn(z)X_{\rm{tot}} qa\\0\end{matrix}\right)\;,\\\mathcal{P}_{\rm{mag}}&=\mathcal{P}_0\left(\begin{matrix}0\\-2sgn(z)X_{\rm{mag}} qa\\0\end{matrix}\right)\;.
\end{align}
From the different longitudinal component of $\mathcal{P}_{\rm{tot}}$ and $\mathcal{P}_{\rm{mag}}$, we infer that the reflection properties of the optical and acoustic mode must be fundamentally different. In the case of the acoustic mode, the chiral nature of the plasmon should be enhanced and show unique (quite likely circular) features in typical SNOM-experiments such as the ones of Ref. \onlinecite{Hesp21}.

\subsection{Chiral resonance}
From the definition of $X_{\rm{tot}}$, we infer that there is a diverging regime for $\chi_{\rm{tot}}=0$. This regime seems to be necessarily realized at the neutrality point for $\omega\to0$, since the total Drude weight has to vanish, $D_{\rm{tot}}=0$. However, in the d.c. limit also the chiral Drude weight needs to vanish, $D_{\rm{chi}}=0$, again due to gauge invariance.\cite{Stauber20} At the neutrality point, no deflection is thus expected even for the acoustic mode. At finite chemical potential, though, Bloch electrons are deviated without a magnetic field as has recently been discussed by several authors.\cite{Stauber18,Stauber18b,Bahamon20,HeWenYo20,Polshyn20,Sharpe21,Antebi22}

At finite frequencies, we expect sweet spots whenever $\chi_{\nu}(\omega)\to0$ with $\nu=\rm{tot},\rm{mag}$. These frequencies lead to $X_{\nu}\to\infty$ which we will denominate as {\it chiral resonances}. At these frequencies, also the plasmonic modes seem to eventually disappear, see Eqs. (\ref{PlasmonDispersionOp}) and (\ref{PlasmonDispersionAc}). However, a coupling between the optical and acoustic mode will emerge and the plasmon dispersion will then also depend on $\chi_{\rm{chi}}$.\cite{,Lin20,Margetis21}

Chiral resonances also occur if $\chi_{\rm{chi}}\gg1$. This is, e.g., the case for a twist angle $\theta=1.3^\circ$ at $\hbar\omega\approx50$meV where $X_{\rm{mag}}\approx10$, see left panel of Fig. \ref{Plasmon}. At this frequency, the Poynting vector is largely enhanced at small wave numbers. Other sweet spots may be limited to low temperatures, e.g., for twist angle $\theta=1.2^\circ$ at $\hbar\omega\approx22$meV, see central panel of Fig. \ref{Plasmon}. At these chiral resonance, we also assume a coupling between the optical and acoustic mode.
  
\subsection{Chiral plasmons at the neutrality point}
A Dirac system does not host plasmons at the neutrality point. Even though electron-hole transitions may lead to positive and negative charge densities, the charge response is always negative such that the RPA-condition for plasmonic excitations can never be fulfilled. 

This changes in moir\'e systems, where flat bands emerge. The moir\'e potential that confines the electrons in the AA-stacked region then acts as restoring force such that the electronic and hole charge density can oscillate in-phase. From a technical point of view, this can be deduced from the highly peaked absorption due to the flat bands as this may lead to a positive charge response due to the Kramer-Kronig relation. As the collective motion is composed of localized electrons, also the plasmonic bands are usually flat.\cite{Stauber16,Levitov19,Novelli20,Hesp21,Kuang21,Huang22} 

One crucial condition for long-lived plasmons is the presence of an optical gap which emerges in the continuum model by considering relaxation effects.\cite{Koshino19} Now, if the absorption is sufficiently peaked, a positive reactive part of the charge excitations can leak inside the optical gap even though there are no nominal charges in the system. This implies the possibility of a mode ("plasmon") as a pole in RPA response. The resulting response functions are shown in Fig. \ref{Plasmon} for different twist angles and temperatures with $\kappa=0.8$.

The features of the plasmonic excitations can be summarized as follows: (i) Optical plasmons can exist right above the optical gap and persist for temperatures up to $T\sim50$K for $\theta<1.3^\circ$. This is similar to the optical plasmons in flat bands with excess charge.\cite{Levitov19,Khaliji20}  (ii) Acoustic plasmons can exist almost in the whole optical window. Most notably, the magnetic Drude weight carries by far the largest optical weight and we expect excitations with frequencies larger than that of the corresponding optical plasmon for $qa\lesssim100$. At the chiral resonance for which $X_{\rm{mag}}(\omega)$ reaches a maximum, these modes are characterized by a largely enhanced energy density $w\sim X_{\rm{mag}}^2$ as can be deduced from the continuity equation and Eq. (\ref{Pacc}).

Let us finally highlight that both plasmon modes are intrinsically chiral since $\chi_{\rm{chi}}$ is finite throughout the protected window. This is due to the broken particle-hole symmetry as will be discussed in the Sec. \ref{SecChiral}.

\section{Chiral response at the neutrality point}
\label{SecChiral}
Chiral response in twisted bilayer graphene has been observed experimentally in Ref. \onlinecite{Kim16} and is thus manifested in misaligned van-der-Waals heterostructures. In Ref. \onlinecite{Suarez17}, it was shown that neglecting the relative rotation of the pseudospin-orientation between the two layers renders the chiral response. The difference in pseudospin orientation, which is a consequence of the real space chiral symmetry, is thus responsible for the chiral response in the non-interacting continuum model. 

In this section, we will directly link the chiral response to particle-hole symmetry and argue how a slight particle-hole asymmetry will lead to a finite chiral response characterized by van Hove singularities. Our results should also be interesting in view of
other mechanisms causing particle-hole breaking, such as non-local tunneling\cite{XieMing21} or Hartree(-Fock) renormalization\cite{Bultinck20,YiZhang20,BiaoLian21,Bernevig21,FangXie21a,Rademaker19,Seo19,Gonzalez21,XieMing21} of the bands.
\subsection{Symmetries of response functions}
The continuum model displays particle-hole symmetry if the pseudo-spin rotation is neglected $\bm \tau_\alpha^\gamma\rightarrow\bm \tau_\alpha$.\cite{Moon13} This can be seen by the following anti-unitary transformation $\mathcal{U}=\mathcal{S}\mathcal{P}\mathcal{K}$. The unitary operator $\mathcal{S}$ reverts the sign of $k_x$, $\mathcal{S}|k_x,k_y,\alpha,\ell\rangle=|-k_x,k_y,\alpha,\ell\rangle$, the unitary operator $\mathcal{P}$ adds a $\pi$-phase to states in layer $\ell=2$, $\mathcal{P}|k_x,k_y,\alpha,2\rangle=-|k_x,k_y,\alpha,2\rangle$, and the complex-conjugate $\mathcal{K}$ effectively changes the sign of $k_y$. We thus have $\mathcal{U}\mathcal{H}\mathcal{U}^{-1}=-\mathcal{H}$.

We can now discuss the effect of $\mathcal{U}$ on the general response function. For this, we suppress the index $\k$ and write
\begin{equation}\label{Kubo}
\chi_{\mathcal{A}\mathcal{B}} = \sum_{n,m}
\frac{n_F(\epsilon_{m}) - n_F(\epsilon_{n})}{\omega+i0^+ - \epsilon_{n} + \epsilon_{m}} 
\langle m|\mathcal{A}|n\rangle \langle n|\mathcal{B}|m\rangle\;.
\end{equation}

Using the eigenbasis $\{\tilde n\}$ of $\mathcal{H}$, with $|\tilde n\rangle=\mathcal{U}|n\rangle$ and $\mathcal{H}|\tilde n\rangle=\epsilon_{\tn}|\tilde n\rangle$ where $\epsilon_{\tn}=-\epsilon_{n}$, one can calculate any response as 
\begin{equation}\label{Kubo}
 \chi_{\mathcal{A}\mathcal{B}} = \sum_{\tn,\tm}
\frac{n_F(\epsilon_{\tm}) - n_F(\epsilon_{\tn})}{\omega+i0^+ - \epsilon_{\tn} + \epsilon_{\tm}} 
\langle \tm|\mathcal{A}|\tn\rangle \langle \tn|\mathcal{B}|\tm\rangle\;.
\end{equation}
We can then write
\begin{align}
\label{DetailedBalance}
\frac{n_F(\epsilon_{\tm}-\mu) - n_F(\epsilon_{\tn}-\mu)}{\omega+i0^+ - \epsilon_{\tn} + \epsilon_{\tm}}=\frac{n_F(\epsilon_{n}+\mu) - n_F(\epsilon_{m}+\mu)}{\omega+i0^+ - \epsilon_{m} + \epsilon_{n}}\;,
\end{align}
where we have explicitly included the chemical potential $\mu$ in the argument of the Fermi function. We now have for the antiunitary transformation $\langle\tn|\phi\rangle=[\langle n|(\mathcal{U}^\dagger|\phi\rangle]^*$. Therefore, we have $\langle \tm|\mathcal{A}|\tn\rangle=\langle m|\tilde{\mathcal{A}}|n\rangle^*=\langle n|\tilde{\mathcal{A}}|m\rangle$ with $\tilde{\mathcal{A}}$ defined below. The particle-hole symmetry $\mathcal{U}$ thus leads to the following relation:
\begin{align}
\chi_{\mathcal{A}\mathcal{B}}(\mu)=\chi_{\tilde{\mathcal{A}}\tilde{\mathcal{B}}}(-\mu)\;,
\end{align}
with $\tilde{\mathcal{A}}=\mathcal{U}\mathcal{A}\mathcal{U}^{-1}$ and $\tilde{\mathcal{B}}=\mathcal{U}\mathcal{B}\mathcal{U}^{-1}$. We now see, because of ${\bm \tau}_x=\tilde{\bm \tau}_x$ and ${\bm \tau}_y=-\tilde{\bm \tau}_y$, that the response obeys the following relations:
\begin{align}
\sigma_0(\mu)&=\sigma_0(-\mu)\\\;
\sigma_1(\mu)&=\sigma_1(-\mu)\\\;
\sigma_{xy}(\mu)&=-\sigma_{xy}(-\mu)\;.
\end{align}
For $\mu=0$, we thus have $\sigma_{xy}=0$ for all temperatures and frequencies as claimed.
\subsection{Electron and hole transitions}
To make the discussion more illustrative, we switch to the particle-hole picture by defining $\epsilon_{n}^e=\epsilon_{n}$ if $\epsilon_{n}>0$ and $\epsilon_{n}^h=-\epsilon_{n}$ if $\epsilon_{n,\bm k}<0$. We only consider vertical transitions and a general transition $n\to m$ at half-filling with $\mu=0$ is now characterized by the initial and final energies, $\epsilon_n^h\to\epsilon_m^e$. 

For the electron-hole symmetric model, there are transitions with $\epsilon_n^h=\epsilon_n^e$. However, this symmetry is usually slightly broken and generally one finds $\epsilon_n^h\neq\epsilon_m^e$. We can thus classify all (relevant) transitions by  either {\em electron} transitions if $\epsilon_{n}^e>\epsilon_{m}^h$ or by {\em hole} transitions if $\epsilon_{n}^e<\epsilon_{m}^h$. 

Let us now denote response functions consisting of electronic (hole) transitions as $\chi^{e(h)}$. The particle-hole transformation $\mathcal{U}$ further relates $\epsilon_{\tn}^e=\epsilon_{\tn}=-\epsilon_{n}=\epsilon_{n}^h$ and $\epsilon_{\tm}^h=-\epsilon_{\tm}=\epsilon_{m}=\epsilon_{m}^e$. We now see, because of ${\bm \tau}_x=\tilde{\bm \tau}_x$ and ${\bm \tau}_y=-\tilde{\bm \tau}_y$, that the response of {\em electron} transitions and {\em hole} transitions obeys the following relations:
\begin{align}
\chi_{xx}^{e}=\chi_{xx}^{h}\;,\;\chi_{xy}^{e}=-\chi_{xy}^{h}
\end{align}

\begin{figure*}[t]
\includegraphics[width=0.72\columnwidth]{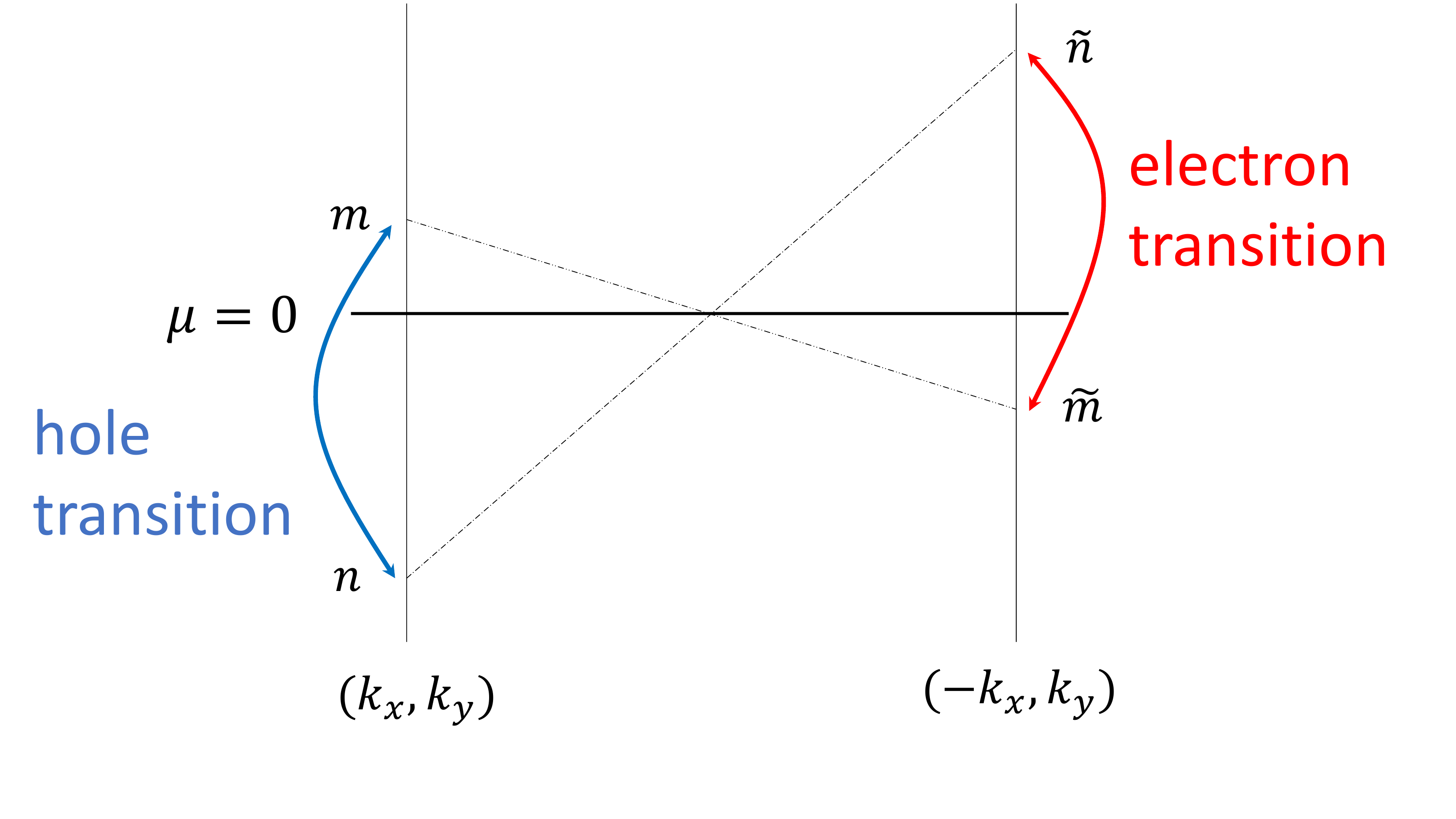}
\includegraphics[width=0.65\columnwidth]{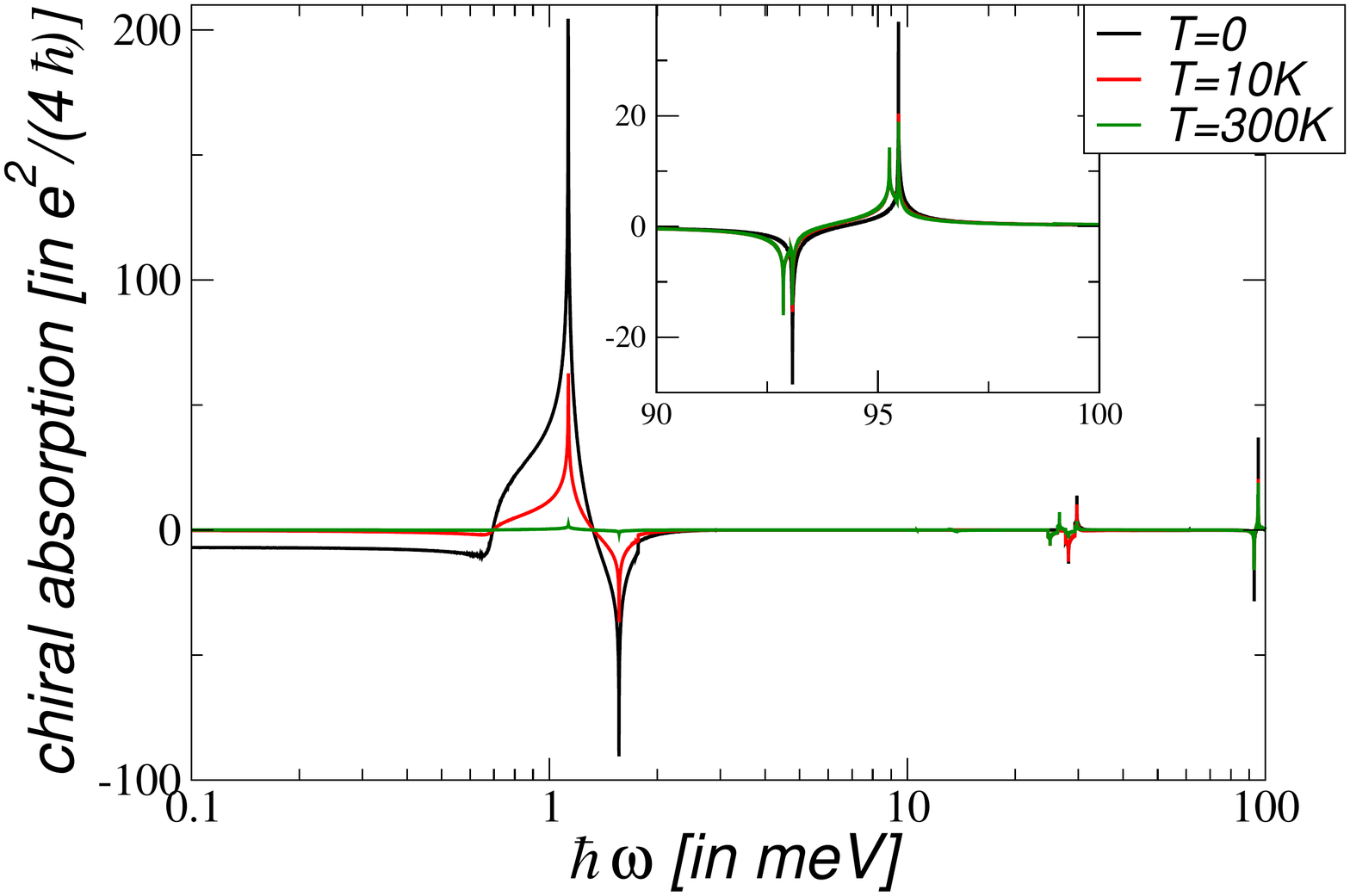}
\includegraphics[width=0.65\columnwidth]{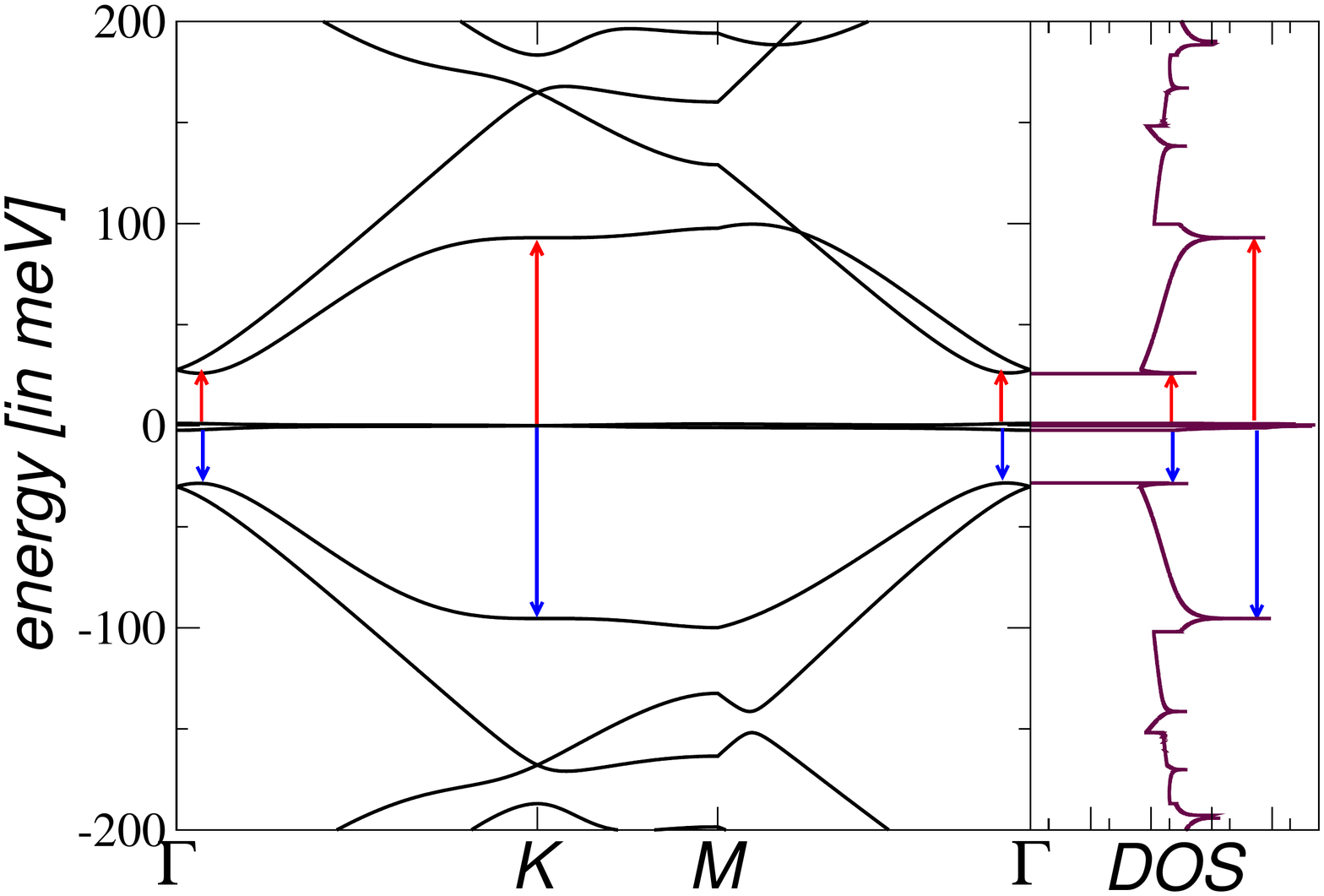}
\caption{\label{FigChiralResponse} Left panel: Illustration of the detailed balance relation of a particle-hole symmtric mode. Via the anti-unitary transformation $\mathcal{U}$, the transitions from $\epsilon_n\to\epsilon_m$ at momentum $(k_x,k_y)$ are directly related to the transitions from $\epsilon_{\tilde m}\to\epsilon_{\tilde n}$ at momentum $(-k_x,k_y)$.  Any hole transition ($\epsilon_{n}^h>\epsilon_{m}^e$) is automatically related to an electron transition ($\epsilon_{\tn}^e>\epsilon_{\tm}^h$) since $\epsilon_{\tn}^e=\epsilon_{n}^h$ and $\epsilon_{\tm}^h=\epsilon_{m}^e$. Center panel: Chiral response Re$\sigma_{\rm{chi}}(\omega)$ of the asymmetric continuum model with $\kappa=0.8$ in Eq. (\ref{Hamiltonian}) at the neutrality point with twist angle $\theta=1.1^\circ$ for temperatures $T=0,10,300$K. The inset highlights the chiral response around $\hbar\omega=95$meV. Right panel: Corresponding band structure and density of states (DOS) on logarithmic scale. The transitions related to the van Hove singularities around $\hbar\omega=25$meV and $\hbar\omega=95$meV are indicated by red (electronic transition) and blue (hole-like transition) arrows.}
\end{figure*}

Numerically, we find that the dominant chiral electron (hole) transitions between different bands and with small energy denominator are negative (positive). However, for larger energy denominators, we also find chiral electronic (hole) transitions which have the opposite sign. Furthermore, the sign of the chiral response due to electron (hole) transitions between the same bands can change. The momenta of electron and hole transitions then normally form a well-defined boundary in the Brillouin-zone. For transitions within the flat bands, however, we also found fractal boundaries. 

\subsection{Detailed balance}
The transformation $\mathcal{U}$ links the momentum $(k_x,k_y)$ to momentum $(-k_x,k_y)$. Eq. (\ref{DetailedBalance}) guarantees that the transition $n\to m$ at momentum $(k_x,k_y)$ from $\epsilon_n$ to $\epsilon_m$ and at chemical potential $-\mu$ carries the same weight as the transition $\tilde m\to\tilde n$ at momentum $-(k_x,k_y)$ from $\epsilon_\tm$ to $\epsilon_\tn$ and at chemical potential $\mu$. Since also the matrix elements have the same (absolute) value, we thus obtain a detailed balance relation for the above transitions at the neutrality point $\mu=0$. This is illustrated in the left panel of Fig. \ref{FigChiralResponse}. 

With $\chi_{\alpha\beta}=\sum_{m,n;k_x,k_y}\chi_{\alpha\beta}(m,n;k_x,k_y)$, we can link a single electron transition to a single hole transition as follows:
\begin{align}
\chi_{xx}(m,n;k_x,k_y)&=\chi_{xx}(n,m;-k_x,k_y)\;,\\
\chi_{xy}(m,n;k_x,k_y)&=-\chi_{xy}(n,m;-k_x,k_y)\;.
\end{align}
This detailed balance between the electron transition at $(k_x,k_y)$ and the corresponding hole transition at $(-k_x,k_y)$ eventually leads to a vanishing chiral response at half-filling.

We can also define a different particle-hole transition as was proposed by Moon and Koshino.\cite{Moon13} Together with time-reversal and rotational symmetry, this leads to 
\begin{align}
\chi_{xx}(m,n;k_x,k_y)&=\chi_{xx}(n,m;-k_x,-k_y)\;,\\
\chi_{xy}(m,n;k_x,k_y)&=-\chi_{xy}(n,m;-k_x,-k_y)\;.
\end{align}

\subsection{Dissipative chiral response close to the magic angle}
We will now discuss the chiral response of the full model of Eq. (\ref{Hamiltonian}) at the neutrality point. Crucially, the rotation in pseudospin-space needs to be included to break particle-hole symmetry as discussed before. However, the approximate electron-hole symmetry suggested by $\mathcal{U}$ will still relate sublattice and layer, leading to a coherence of the wave function between these two degrees of freedom which must not be related to the underlying lattice (spatial) symmetry.\cite{Stauber20C,Ochoa20}  

Since electron-hole symmetry is slightly broken, we can label all transitions as either electron or hole transitions. The electronic wave function is not strongly affected by this small perturbation and due to continuity arguments, around certain regions in $\k$-space, electron and hole transitions must still have well-defined, but opposite signs.

Apart from the transition-matrix element, the response is also determined by the electronic dispersion. In any Bloch-band, there is at least one van Hove singularity and in principle, we expect an enhanced optical response if either the initial or final state is located at one singular $\k$-point. However, the transition-matrix element might be suppressed due to symmetries and precisely the approximate particle-hole symmetry suppresses the optical transitions of the total current at the $M$-point.\cite{Moon13} This is not the case, though, for the magnetic and chiral transitions and we thus expect a large response due to the large van Hove singularity which can also be located around the $K$ or $\Gamma$-point.

In the electron-hole symmetric model, van Hove singularities necessarily appear in the occupied and unoccupied bands at $\epsilon_{vH}^h=\epsilon_{vH}^e$. Slightly breaking this symmetry will lead to a splitting with $\epsilon_{vH}^h\neq\epsilon_{vH}^e$. Possible transitions are now of electron and hole nature that have opposite chiral response, but do not cancel each other anymore. Also the band-edges of the electronic and hole bands will slightly shift due to the broken symmetry, given rise to either pure electron or hole transitions. To conclude, we expect prominent features coming from singularities of the band structure, either discontinuities or logarithmic divergencies, where the electronic and hole transitions are not compensated by each other.

This can be seen in the center panel of Fig. \ref{FigChiralResponse} where the dissipative response of twisted bilayer with twist angle $\theta=1.1^\circ$ and $\kappa=0.8$ is shown. There are always two peaks that come in pairs, a negative peak and a positive peak associated with either electron or hole transitions. 

The first pair originates from transitions within the flat bands and is strongly temperature dependent, i.e., practicable absent at room temperature. The second and third pair are related to transitions from the flat to the first remote band and associated to van Hove singularities located at the $\Gamma$ and $K$-point, respectively. They thus do not as strongly depend on temperature and in both cases, the negative (positive) response is related to electron (hole) transitions. The response of the third pair is highlighted in the inset of the center panel of Fig. \ref{FigChiralResponse} for the sake of clarity.

In the right panel of Fig. \ref{FigChiralResponse}, the band structure is shown and the electron (red arrow) and hole (blue arrow) transitions are shown for the second and third pairs. Generally, we expect strong chiral response at energies involving a large density of states. These energies can be identified from the density-of-states (DOS), shown next to the band structure. However, the larger the transition energy becomes, the weaker the response is. 

\section{Summary and Outlook}
Technically speaking, we have investigated the full optical response of magic angle graphene at the neutrality point consisting of the total, magnetic and chiral response. The dissipative response is obtained without the usual damping term by analytically integrating the delta-function on a linearized grid. The reactive response is then obtained via the Kramers-Kronig relation applying a suitable cutoff for large frequencies. By this, we obtain accurate results close to the magic angle even for low energies.

Generally speaking, we have investigated the continuum model introduced in Refs. \onlinecite{Lopes07,Bistritzer11} which resembles the standard model to address general topics related to van-der-Waals heterostructures. This model is believed to be well-understood, but here we showed that the ground-state of the {\it non-interacting} continuum model at the neutrality point is unstable in the immediate vicinity of the magic angle with respect to transverse current fluctuations. We thus predict a so-called Condon instability\cite{Guerci21} using a novel scaling approach. 

The Condon instability at the magic angle is supposedly interesting only from a theoretical point of view. However, we also presented new results with high potential for technological impact. We pointed out that the plasmonic bonding mode (acoustic or magnetic plasmon) should be larger in energy than the ordinary plasmonic anti-bonding mode (optical or electric plasmon). Furthermore, the energy density of this acoustic mode can be largely enhanced at a certain frequency which we label as {\it chiral} resonance. This novel resonance has not been discussed in the literature so far and should be present for a wide range of twist angles and temperatures.

Another interesting aspect concerns an effective model to describe the physics around the magic angle,\cite{Hejazi19} initially proposed in Refs. \onlinecite{Bena11,Montambaux12} in a different context. This model makes use of an effective parameter that stands for the twist angle, and we now provided a direct mapping to the standard continuum model of twisted bilayer graphene, i.e., to the real twist angle. We also included a momentum-dependent mass-term that makes sure that the universal conductivity of $\frac{e^2}{2\hbar}$ is reached for $\omega\to0$.

Lastly, we discussed the chiral aspects of the continuum model and outlined in detail the implications of an approximate particle-hole symmetry. We distinguished between electron and hole transitions that give equal contributions to the chiral response, but which cancel exactly. Since particle-hole symmetry is generally broken, we show that the {\it finite} chiral response usually comes in pairs consisting of a positive and negative signal since electron and hole transitions have opposite chirality, respectively.

To conclude, we hope that our results on the Condon instability will stimulate new analytical studies of the continuum model at the magic angle regime. We further hope that our results on the acoustic plasmonic excitations with its chiral features will stimulate experiments which pave the way towards technological use of this phenomenon. 
\section{Acknowledgments}
This work was supported by the mobility program Salvador Madariaga under PRX19/00024 and by the projects No. PGC2018-096955-B-C42, No. PID2020-113164GB-i00, and No. CEX2018-000805-M financed by MCIN/ AEI/10.13039/501100011033. The access to computational resources of CESGA (Centro de Supercomputaci\'on de Galicia) is also gratefully acknowledged. The work of T.S. and of J.S. was further supported by Deutsche Forschungsgemeinschaft via SFB 1277. D.M. wishes to thank Dr. A. B. Watson for an inspiring discussion on the twisted bilayer graphene near the magic angle.

\appendix
 \section{Numerical integration of a generalized density of states}
\label{AppDeltaIntegration}
In this appendix, we describe the numerical recipe how to obtain the optical response functions without introducing the usual damping term. The main numerical task in our approach is the numerical evaluation of two-dimensional integrals that involve a delta-function. If we determine this integral up to large frequencies $\omega$, we can take advantage of the Kramers-Kronig relation in order to obtain the reactive part of the response function. The remaining one-dimensional integral over frequencies does usually not pose any difficulties and the recipe concerning the cut-off procedure has been outlined in Ref. \onlinecite{Stauber13}.

\begin{figure}[t]
\includegraphics[width=0.99\columnwidth]{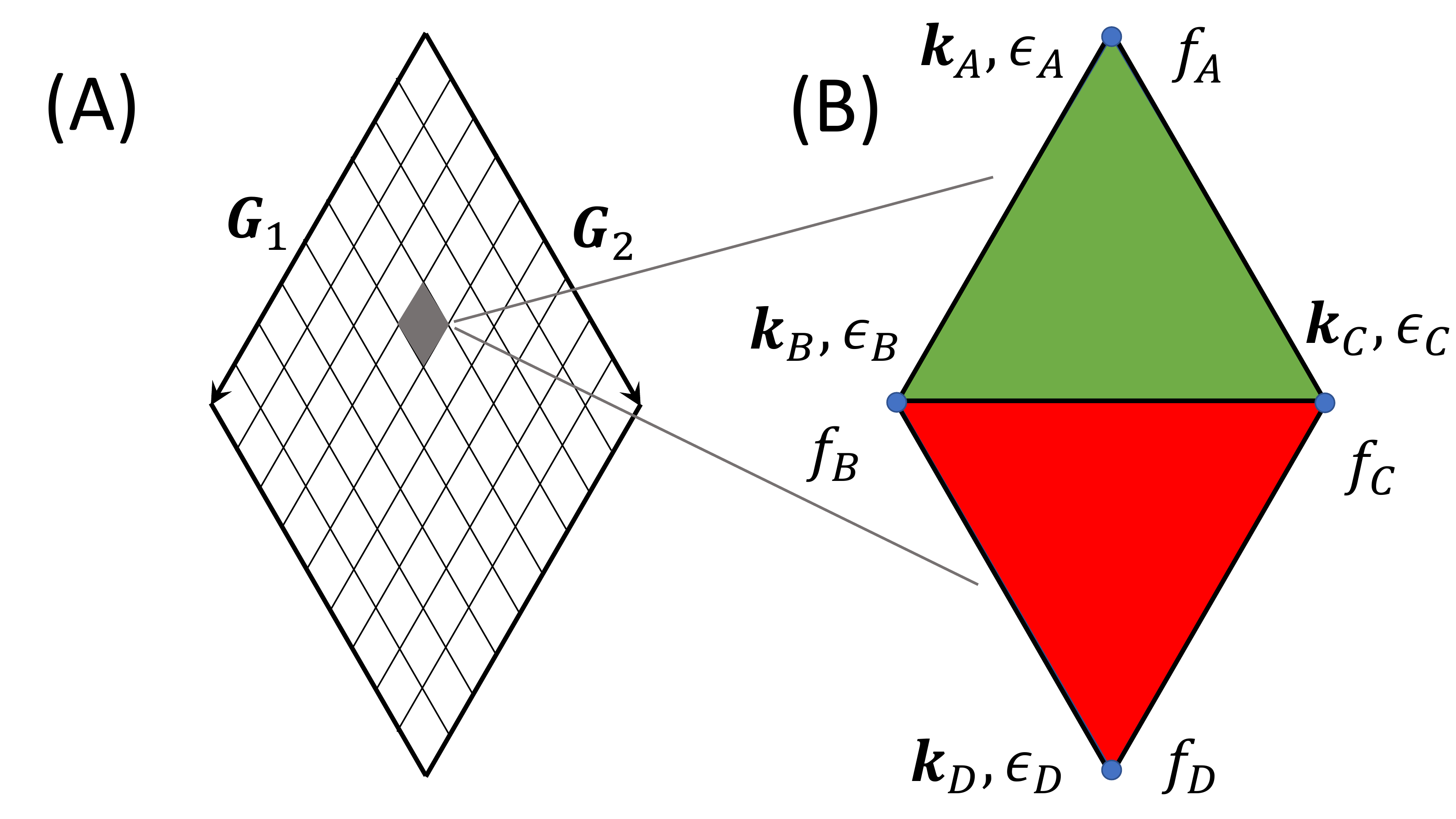}
\caption{\label{Brillouin} (A) The rhombic Brillouin zone defined by the reciprocal lattice vectors ${\bm G}_1$ and ${\bm G}_2$. (B) Zoom-In of a small rhombus with side length $|{\bm G}_1|/N$. The vertices are labeled by $i=A,B,C,D$ and characterized by $\k_i$, $\epsilon_i$, and optionally also by $f_i$.}
\end{figure}

We will calculate the response {\it without} disorder, i.e., we will take the delta-function literally and perform the integration {\it analytically} after having discretized the Brillouin zone's . This can be done by introducing a triangular grid on the Brillouin zone and assuming a linear interpolation. In Fig. \ref{Brillouin}, we show the discretization used for the calculations. We have checked that the final result does not crucially depend on the discretization. Another optimization is obtained by assuming a quadratic interpolation between the three base-points.\cite{Wiesenekker88} We have checked that for large grids used here, this does not lead to significant improvements, also nicely explained in Ref. \onlinecite{Perdersen08}. 

We discretize the Brillouin zone by $N$ with $n,m=0,...,N$ in the following way:
\begin{align}
\k=\frac{n}{N}{\bm G}_1+\frac{m}{N}{\bm G}_2\;,
\end{align}
with the lattice vectors  $\bm G_1 =  |\Delta \bm K| \left( \tfrac{-\sqrt{3}}{2},-\tfrac{3}{2}\right)$, $\bm G_2 =  |\Delta \bm K| \left(\tfrac{\sqrt{3}}{2},-\tfrac{3}{2}\right)$, see Fig. \ref{Brillouin} A). In our calculations, we chose discretizations up to $N\cong10000$; for twist angles in the immediate vicinity of the magic angle even as large as $N\cong20000$.

We shall calculate the following generalized density of states with $g$ denoting a degeneracy factor:
\begin{align}
\rho(\epsilon)=\frac{g}{\mathcal{A}}\sum_\k f_\k\delta(\epsilon-\epsilon_\k)
\end{align}
As we assume periodic boundary conditions, the sample area is given by $\mathcal{A}=N^2A_c$ where $A_c$ is the area of the unit cell. In the case of twisted bilayer graphene, we have $A_c=\frac{\sqrt{3}}{2}a_g^2A_i$ as the area of the moir\'e supercell with $a_g=2.46 \, \mathring{\text{A}}$, $A_i=3i^2 + 3i + 1$ and $\cos(\theta_i)=1-\tfrac{1}{2A_i}$.  

We will now consider each of the $N^2$ mini-rhombi individually which are characterized by the vertices $\k_i$, $\epsilon_i$, and optionally $f_i$ with $i=A,B,C,D$. First, we will divide the mini-rombus in two and consider first the triangle defined by $i=A,B,C$ and afterwards the triangle defined by $i=B,C,D$. 

To outline the algorithm, we will only consider the first triangle and further assume that $\epsilon_A\leq\epsilon_B\leq\epsilon_C$ which can always be achieved by relabelling the vertices. We now interpolate linearly between the three vertices such that any momentum $\k$ and energy $\epsilon$ inside the triangle can be parameterized by two parameters $t,s\in[0,1]$ (due to the prior ordering):
\begin{widetext}
\begin{align}
\left(
\begin{matrix}
k_x\\k_y\\\epsilon_\k
\end{matrix}
\right)=
\left(
\begin{matrix}
k_{B,x}-k_{A,x}\\k_{B,y}-k_{A,y}\\\epsilon_{\k_B}-\epsilon_{\k_A}
\end{matrix}
\right)t+
\left(
\begin{matrix}
k_{C,x}-k_{A,x}\\k_{C,y}-k_{A,y}\\\epsilon_{\k_C}-\epsilon_{\k_A}
\end{matrix}
\right)s+
\left(
\begin{matrix}
k_{A,x}\\k_{A,y}\\\epsilon_{\k_A}
\end{matrix}
\right)\;,
\end{align}
\end{widetext}
We can now write the integral that contains the contribution $\rho_\Delta$ to $\rho$ over the triangle with respect to the two variables $t$ and $s$. The integration limits corresponding to the vertices $[A,B,C]$ are now given with respect to the axis defined by $t,s$, i.e., $[(0,0),(1,0),(0,1)]$. Neglecting for the moment the weight function $f_\k$ and setting $g=1$, we arrive at the following expression:
\begin{align}
\rho_\triangle(\epsilon)&=\frac{1}{(2\pi)^2} \int_\triangle d^2k\;\delta(\epsilon-\epsilon_\k)\;,\\
&=\frac{\tilde J}{(2\pi)^2} \int_0^1dt\int_0^{1-t}ds\;\delta(s-s(\epsilon,t))\;, 
\end{align}
where we introduced the Jacobian $J=|(k_{B,x}-k_{A,x})(k_{C,y}-k_{A,y})-(k_{C,x}-k_{A,x})(k_{B,y}-k_{A,y})|$ with $\tilde J=J/(\epsilon_C-\epsilon_A)$ and $s(\epsilon,t)=\frac{\epsilon-\epsilon_A}{\epsilon_C-\epsilon_A}-\frac{\epsilon_B-\epsilon_A}{\epsilon_C-\epsilon_A}t$. 

The integral depends on the value of $\epsilon$ relative to the energies $\epsilon_i$ and we obtain
\begin{align}
\rho_\Delta(\epsilon)=\frac{\tilde J}{(2\pi)^2}
\Big[&\frac{\epsilon-\epsilon_A}{\epsilon_B-\epsilon_A}\theta(\epsilon-\epsilon_A)\theta(\epsilon_B-\epsilon)\notag\\
+&\frac{\epsilon_C-\epsilon}{\epsilon_C-\epsilon_B}\theta(\epsilon-\epsilon_B)\theta(\epsilon_C-\epsilon)\Big]\;.
\end{align}
The total density of states is then obtained by the sum $\rho(\epsilon)=\sum_\Delta\rho_\Delta(\epsilon)$.

The weight function $f_\k$ can now be included by linear interpolation.  With
\begin{align}
f(\epsilon)=&f_B\frac{\epsilon-\epsilon_A}{\epsilon_B-\epsilon_A}\theta(\epsilon-\epsilon_A)\theta(\epsilon_B-\epsilon)\notag\\
+&f_A\frac{\epsilon_B-\epsilon}{\epsilon_B-\epsilon_A}\theta(\epsilon-\epsilon_A)\theta(\epsilon_B-\epsilon)\notag\\
+&f_C\frac{\epsilon-\epsilon_B}{\epsilon_C-\epsilon_B}\theta(\epsilon-\epsilon_B)\theta(\epsilon_C-\epsilon)\notag\\
+&f_B\frac{\epsilon_C-\epsilon}{\epsilon_C-\epsilon_B}\theta(\epsilon-\epsilon_B)\theta(\epsilon_C-\epsilon)\;,
\end{align}
and reincorporation of the degeneracy factor $g$, the generalized density of states is thus approximated by
\begin{align}
\rho_\Delta(\epsilon)=\frac{g\tilde J}{(2\pi)^2}f(\epsilon)
\Big[&\frac{\epsilon-\epsilon_A}{\epsilon_B-\epsilon_A}\theta(\epsilon-\epsilon_A)\theta(\epsilon_B-\epsilon)\notag\\
+&\frac{\epsilon_C-\epsilon}{\epsilon_C-\epsilon_B}\theta(\epsilon-\epsilon_B)\theta(\epsilon_C-\epsilon)\Big]\;.
\end{align}
Apart from increasing the discretization, the numerical results can be further smoothened by explicitly taking advantage of the rotational symmetry, i.e., $2\j_{\rm{tot}}\cdot\j_{\rm{tot}}=(j_x^1+j_x^2)^2+(j_y^1+j_y^2)^2$, $2\j_{\rm{mag}}\cdot\j_{\rm{mag}}=(j_x^1-j_x^2)^2+(j_y^1-j_y^2)^2$, and $2\j_{xy}\cdot\j_{xy}=j_x^1j_y^2-j_x^2j_y^1$.

\section{Real part of interband conductivity: Analytical derivations}

In this appendix, we describe analytically the real part of the interband conductivity for $\mu=0$ and $T=0$, by use of the two-band model introduced in Refs. \onlinecite{Bena11,Montambaux12}. In this model, the Dirac cone coexists with a parabolic profile in the Hamiltonian. We focus on the limits of the interband conductivity as $\omega\to 0$ and $\omega\to\infty$.

\subsection{Model Hamiltonian}
\label{subsec:formalism}
The reduced, two-band Hamiltonian without a gap reads
\begin{equation}
H_{red}=-	
\begin{pmatrix}
0 & \varpi^2+\eta \varpi^*	\\
{\varpi^*}^2+\eta \varpi &  0 
\end{pmatrix}~,
\end{equation}
where $\varpi=k_x\mp i k_y$ in the vicinity of $\boldsymbol K$ ($\boldsymbol K'$). Here, we have set $\hbar=1=2m$ for later algebraic convenience. The parameter $\eta$ is assumed positive and small ($0<\eta\ll 1$). It expresses the relative strength of the Dirac cone. From now on, we focus on the point $\boldsymbol K$. We will comment on the case with $\eta<0$ below.

This Hamiltonian yields the eigenenergies
\begin{equation}
\epsilon_{\boldsymbol k,\pm}=\pm |F_{\boldsymbol k}|~,\quad F_{\boldsymbol k}={\varpi^*}^2+\eta \varpi~,
\end{equation}
and the normalized eigenvectors
\begin{equation}
|\pm\rangle_{\boldsymbol k}=	\frac{1}{\sqrt{2}} 
\begin{pmatrix} 1 \\
                \mp e^{i \vartheta_{\boldsymbol k}}	
\end{pmatrix}~,\quad 
 \vartheta_{\boldsymbol k}={\rm Arg}F_{\boldsymbol k}~.
\end{equation}
The eigenenergies are expressed explicitly by
\begin{equation}\label{eq:E-q}
\epsilon_{\boldsymbol k, \pm}= \pm\sqrt{k^4+2\eta k^3\cos(3\theta)+\eta^2k^2}=\pm\epsilon(\boldsymbol k;\eta)~,	
\end{equation}
in the polar coordinates $(k, \theta)$ with center at $\boldsymbol K$.

Evidently, the scaling of the momentum with $\eta$ according to $\boldsymbol k =\eta \tilde{\boldsymbol k}$ results in $\epsilon(\boldsymbol k;\eta)=\eta^2 \tilde \epsilon(\tilde{\boldsymbol k})$ where 
\begin{equation}\label{eq:E-q-scaled}
\tilde \epsilon(\boldsymbol k)= \sqrt{k^4+2 k^3\cos(3\theta)+k^2}=\epsilon(\boldsymbol k; 1)~.	
\end{equation}
It is algebraically convenient to use the scaled momentum and eigenenergy (see, however, Eq.~\eqref{eq:reg_cond-Kubo}). For ease of notation, we henceforth drop the tildes from $\tilde{\boldsymbol k}$ and $\tilde \epsilon$.

Next, we describe the local minima of $\epsilon_{\boldsymbol k, +}=\epsilon(\boldsymbol k)$.
By $\nabla_{\boldsymbol k} (\epsilon^2)=0$ we obtain  $\boldsymbol k=k_c(\cos\theta_c, \sin\theta_c)$ where $k_c=0$, or $k_c=1$ with $\theta_c=\pi - 2\pi n/3$, $n\in \mathbb{N}$. These points yield zero bandgap. The other critical points of $\epsilon(\boldsymbol k)$ correspond to saddle points, with nonzero bandgap, and are disregarded. If $\eta<0$, the local minima correspond to $k_c=0$, or $k_c=1$ with $\theta_c= 2\pi n/3$ (by $k_x \rightarrow -k_x$).

We turn our attention to the velocity matrix element needed for the interband conductivity. By setting $F=F_R+ i F_I$ ($F_R=\Re F$ and $F_I=\Im F$), we have 
\begin{align*}
	\langle -|\nabla_{\boldsymbol k} H_{red} |+\rangle &= -i \epsilon(\boldsymbol k)\,\nabla_{\boldsymbol k}\vartheta_{\boldsymbol k} \notag\\
	&=\frac{i}{\epsilon(\boldsymbol k)}\{(\nabla_{\boldsymbol k} F_R)F_I-F_R(\nabla_{\boldsymbol k} F_I)\}~.
\end{align*}
Let's compute the $x$-component, for example. We find
\begin{equation*}
	\langle -|\partial_{k_x} H_{red} |+\rangle = ~ -2i \, \frac{\Lambda_x(\boldsymbol k)}{\epsilon(\boldsymbol k)} 
\end{equation*}
where
\begin{equation}\label{eq:Lambda-def}
\Lambda_x(\boldsymbol k)=k_y\left\{\left(k_x-\frac{1}{2}\right)^2+k_y^2-\frac{3}{4} \right\}~.
\end{equation}
For $\boldsymbol K'$, one simply has to replace $k_x$ by $-k_x$. Note that $\Lambda_x(\boldsymbol k)=0$ at the local minima of $\epsilon(\boldsymbol k)$ determined above.

\subsection{Integral of interband conductivity}
\label{subsec:interband}
The diagonal elements of the interband (regular) conductivity are computed from the formula ($\alpha = x,\, y$)
\begin{widetext}
\begin{equation}\label{eq:reg_cond-Kubo}
\sigma_{\alpha\alpha}^R(\omega)= 4ig_s g_vg_\ell\sigma_G\ (\omega+i 0^+)\int\limits\frac{d^2{\boldsymbol k}}{(2\pi)^2}\ \frac{n_F(\epsilon(\boldsymbol k;\eta))-n_F(-\epsilon(\boldsymbol k;\eta))}{\epsilon(\boldsymbol k; \eta)}   
\frac{|\langle -| \partial_{k_\alpha}H_{red}|+\rangle|^2}{4\epsilon(\boldsymbol k;\eta)^2-(\omega+i 0^+)^2}~;\quad \sigma_G=\frac{e^2}{4\hbar}.  
\end{equation}
\end{widetext}
Here, the factor $g_s g_vg_\ell$ accounts for the layer-degree of freedom, and the usual spin and valley degeneracies. In Eq.~\eqref{eq:reg_cond-Kubo} we use the unscaled momentum $\boldsymbol k$ and the eigenenergy $\epsilon(\boldsymbol k;\eta)$ from Eq.~\eqref{eq:E-q}. We set $\mu=0$ and $T=0$, take $\alpha=x$, and change the integration variable from $\boldsymbol k$ to $\eta\boldsymbol k$. Thus, we arrive at the simplified integral
\begin{equation}\label{eq:reg_cond-Kubo-simpl}
\sigma_{xx}^R(\omega)= -8i g_s g_vg_\ell\sigma_G \tilde\omega\int\limits\frac{d^2{\boldsymbol k}}{(2\pi)^2}\ \frac{1}{\epsilon(\boldsymbol k)^3}   
\frac{\Lambda_x(\boldsymbol k)^2}{\epsilon(\boldsymbol k)^2-(\tilde \omega+i 0^+)^2}  
\end{equation}
where $\epsilon(\boldsymbol k)$ is given by Eq.~\eqref{eq:E-q-scaled} and 
\begin{equation}\label{eq:omega-scaled}
\tilde\omega = \frac{\omega}{2\eta^2}~.	
\end{equation}
We will keep the symbol $\tilde\omega$ (with tilde) throughout.

Our task is to compute $\Re \sigma_{xx}^R$ by carrying out the integration in \emph{local} polar coordinates by consideration of points $\boldsymbol k=\boldsymbol k_*$ such that $\epsilon(\boldsymbol k)=\tilde \omega$  (if $\tilde \omega>0$). A difficulty is that these points may locally form non-circular curves. The integral for $\Re\sigma_{xx}^R$ has significant contributions from the vicinity of each curve. We study the following limits: (i) $\tilde\omega\to 0$, when the curves of interest are formed near local minima of $\epsilon(\boldsymbol k)$; and (ii) $\tilde\omega\to +\infty$, when $k_*$ is large.  

\subsubsection{Limit $\tilde\omega\to 0$}
\label{sssec:zero-lim}
For each critical point of interest we set $\boldsymbol q=\boldsymbol k- k_c(\cos\theta_c,\sin\theta_c)$, and find a suitable expansion for the solutions $\boldsymbol q$ of $\epsilon(\boldsymbol k)=\tilde \omega$ by perturbations if $\tilde\omega\ll 1$. For this purpose, we invoke the local polar coordinates $(q,\phi)$, where $\boldsymbol q=q(\cos\phi, \sin\phi)$ ($0\le \phi< 2\pi$); and determine $q=|\boldsymbol q|$ as a function of $\phi$ and $\tilde\omega$. Let $q_*(\phi)$ be such a solution. Subsequently, we expand $\epsilon(\boldsymbol k)^2$ near $q=q_*$.

First, consider $k_c=0$, which amounts to the center point ($\boldsymbol K$). After some algebra, we obtain
\begin{equation}
q_*=\tilde\omega \left\{1-\tilde\omega \cos(3\phi)+ O(\tilde\omega^2)\right\}~,	\end{equation}
where $O(\tilde\omega^2)$ denotes a correction of the order of $\tilde\omega^2$. This formula entails an approximation of the form
\begin{align}\label{eq:epsilon-exp}
	\epsilon(\boldsymbol k)^2-\tilde\omega^2 &\simeq Q_1(\phi)\, (q-q_*) + Q_2(\phi)\,(q-q_*)^2
\end{align}
where $Q_1(\phi)=2q_* [ 1+3q_*\cos(3\phi)] $ and $Q_2(\phi)=1$. 

Second, consider $k_c=1$ with $\theta_c=\pi$, which amounts to the critical point at $\boldsymbol k=(-1, 0)$, for $n=0$. We find
\begin{equation}
q_*\simeq \frac{\tilde\omega}{\sqrt{1+8\sin^2\phi}}\left\{1+\tilde\omega \frac{\cos\phi (1+4\sin^2\phi)}{(1+8\sin^2\phi)^{\frac32}} \right\}~.
\end{equation}
This formula implies expansion~\eqref{eq:epsilon-exp} with
\begin{align}
	Q_1(\phi)= 2q_* \left[(1+8\sin^2 \phi)+3q_*(1+4\sin^2\phi)\right] 
\end{align}
and $Q_2(\phi)=1+8\sin^2\phi$.

Third, we consider the critical points with $k_c=1$ and $\theta_c=\pi-2\pi n/3$ for $n=1,\,2$, i.e., at $\boldsymbol k=(1/2, \pm \sqrt{3}/2)$. We thus obtain the following expansions for $q_*=q_*(\phi)$:
\begin{align}
q_* &\simeq \frac{\tilde\omega}{\sqrt{4\cos^2\phi +3\mp 2\sqrt{3}\sin(2\phi)}}\left\{ 1 -\tilde \omega \right. \notag \\
& \qquad \left. \times  \frac{2\cos\phi \cos(2\phi)\pm \sqrt{3} \sin\phi}{[4\cos^2\phi +3 \mp 2\sqrt{3} \sin(2\phi)]^{\frac32}} \right\}~.	
\end{align}
Each of these formulas implies expansion~\eqref{eq:epsilon-exp} with
\begin{widetext}
\begin{equation}
	Q_1(\phi)= 2q_* \left\{4\cos^2\phi+3\mp 2\sqrt{3}\sin(2\phi)+3q_*[2\cos\phi \cos(2\phi)\pm \sqrt{3}\sin\phi] +2q_*^2 \right\}
\end{equation}
\end{widetext}
and $Q_2(\phi)=4\cos^2\phi +3 \mp 2 \sqrt{3} \sin(2\phi)$. 

In all of the above cases, we have $Q_1(\phi)\neq 0$ for every $\phi$. The expansions for $\epsilon(\boldsymbol k)$ near local minima are uniform in $\phi$; and capture the zero bandgap with a negligible correction of the order of $\tilde\omega^{3/2}$ or smaller. This property can be used to show (as a self-consistency check) that our leading-order result for $\Re\sigma_{xx}^R$, given below, has a negligible correction if $\tilde\omega\ll 1$. We omit details on this here.   

Next, by Eq.~\eqref{eq:reg_cond-Kubo-simpl}, we split the integral for $\sigma_{xx}^R(\omega)$  into four contributions, one for each local minimum of $\epsilon(\boldsymbol k)$. Using the local polar coordinates $(q,\phi)$, we first carry out the integration in $q$ by employing the formula
\begin{equation*}
\frac{1}{\epsilon(\boldsymbol k)^2-(\tilde\omega+i 0^+)^2}=i\pi Q_1(\phi)^{-1}\delta(q-q_*(\phi))+\mathcal P\left(\frac{1}{q-q_*} \right)
	\end{equation*}
for each contribution.
In the above, $\mathcal P(\cdot)$ indicates the principal-value integral. The two-dimensional integral for $\Re\sigma_{xx}^R$ immediately reduces to an integral with respect to the polar angle $\phi$, from the delta function term.

Accordingly, we perform the remaining integration, with respect to $\phi$. For $\tilde\omega\ll 1$,  we write
\begin{equation}
	\Re\sigma_{xx}^R(\omega)\simeq \frac{1}{2}g_\ell g_s g_v\sigma_G \left\{ I^{(c)}+9\sum_{n=0}^2 I^{(n)}\right\}~,
\end{equation}
where $I^{(c)}$ and $I^{(n)}$ correspond to the center point ($k_c=0$) and the points $k_c=1$ and $\theta_c=\pi-2\pi n/3$, respectively. We define and compute the following requisite integrals:
\begin{equation*}
	I^{(c)}=\frac{1}{2\pi}\int_0^{2\pi}d\phi \,\sin^2\phi= \frac{1}{2}~,
\end{equation*}
\begin{equation*}
I^{(0)}=\frac{1}{2\pi}\int_0^{2\pi}d\phi\, \frac{\sin^2\phi}{(1+8\sin^2\phi)^2}=\frac{1}{54}~,	
\end{equation*}
\begin{widetext}
\begin{equation*}
I^{(1)}+I^{(2)}=\frac{1}{2\pi}\int_0^{2\pi}d\phi\, \sin^2\phi \sum_{s=\pm}\frac{1}{[4\cos^2\phi+3+s 2\sqrt{3}\sin(2\phi)]^2}=\frac{7}{27}~.	
\end{equation*}
\end{widetext}
Hence, we finally obtain 
\begin{equation}\label{eq:re-interb-cond-zero-omeg}
	\Re\sigma_{xx}^R(\omega)= \frac{3}{2} g_\ell g_s g_v \sigma_G=12 \sigma_G \qquad \mbox{as}\ \tilde\omega\to 0~,
\end{equation}
for the TBG system. The anticipated correction to this result for small nonzero $\tilde\omega$ is of the order of $\tilde\omega^2$.

\subsubsection{Limit $\tilde\omega\to \infty$}
\label{sssec:inf-lim}
In this case, we apply a procedure similar to the above. In particular, we solve the equation $\epsilon(\boldsymbol k)=\tilde\omega$ for large $\tilde\omega$, to find $k=k_*(\phi)\gg 1$. Then we expand the difference $\epsilon(\boldsymbol k)^2-\tilde\omega^2$ in powers of $k-k_*$. The quantity $\Re\sigma_{xx}^R$ is determined by integration near the curve $k=k_*(\phi)$.

In detail, by perturbations for $\tilde\omega\gg 1$, we obtain
\begin{equation}
	k_* = \sqrt{\tilde\omega}\left\{ 1+\frac{1}{2}\tilde\omega^{-1/2}\cos(3\theta)+O(\tilde\omega^{-1}) \right\}~.
\end{equation}
Here, we use the polar angle $\theta$, where $\boldsymbol k= k(\cos\theta, \sin\theta)$.
The above formula implies expansion~\eqref{eq:epsilon-exp} with $q=k$ and $\phi=\theta$ (since $\boldsymbol k=\boldsymbol q$ here), while
\begin{equation}
Q_1(\theta)=2k_* \left[2k_*^2-3k_* \cos(3\theta)+1  \right]	
\end{equation}
and $Q_2(\theta)=6k_*^2-6 k_* \cos(3\theta)+1$. 

These considerations lead to the simplified integral
\begin{align}
	\Re\sigma_{xx}^R(\omega)&\simeq g_\ell g_s g_v \sigma_G \tilde\omega^{-2}\frac{1}{2\pi}\int_0^{2\pi}d\theta\,k_*(\theta)^4 \sin^2\theta \notag \\
	& \simeq \frac{1}{2} g_\ell g_s g_v \sigma_G= 4 \sigma_G\quad \mbox{as}\ \tilde\omega\to +\infty~,
\end{align}
for the TBG system. 
\end{document}